\documentclass[%
 reprint,
 amsmath,amssymb,
 aps,
prb,
]{revtex4-2}

\usepackage[utf8]{inputenc}
\usepackage{graphicx}
\usepackage{overpic}
 \usepackage{rotating}
 \usepackage{mwe, tikz}
 \usepackage{amsmath,amssymb}
\usepackage{pgfplots}
\usetikzlibrary{calc,arrows,decorations.markings,shapes.geometric} %--- draw lines
\usepackage{amssymb}

\usepackage[inline]{trackchanges}

\usepackage{lipsum}
\usepackage{romannum}
\usepackage{xfp}

\soulregister\ref7
\soulregister\cite7
\soulregister\citeA7
\soulregister\romannum7

\usepackage{graphicx}
\usepackage{amsmath}
\usepackage{mathtools} %matrix
\usepackage{color}
\usepackage{overpic}
\usepackage[export]{adjustbox}
\usepackage{mwe, tikz}
\usepackage{rotating}
\usepackage{latexsym}
\usepackage{csquotes}
\usepackage{tikz} %--- draw lines
\usetikzlibrary{calc,arrows,decorations.markings} %--- draw lines
\usepackage{pict2e}
\usepackage{pgfplots}
\usepackage{float}
\usepackage{hyperref}
\usepackage{xr}
\usepackage{rotating}
\usepackage{amsmath,amssymb}
\usepackage{fp}
\usetikzlibrary{calc} 
\usepackage{booktabs, tabularx}
\usepackage{romannum}
\usepackage{bm}
\usepackage{float}
\usepackage{adjustbox}
\usepackage{booktabs, tabularx}
\usepackage{longtable}

\definecolor{light-gray}{gray}{0.85}

\newcommand{\drawSlope}[6]{ %(xc,yc,a,theta,color,text)
				\coordinate (center1) at (#1,#2); % center of figure
				 \FPeval{\nx}{cos(#4*pi/180)}%
				 \FPeval{\ny}{sin(#4*pi/180)}%
				 \FPeval{\absNx}{abs(\nx)}%
				 \FPeval{\absNy}{abs(\ny)}%
				\coordinate (b) at ($ (center1) + #3*(-\ny,+\nx) $);
				\coordinate (c) at ($ (center1) - #3*(-\ny,+\nx) $); 
				 \FPeval{\xx}{(-\nx*\ny)}%
				\ifdim\xx pt < 0pt 
				\coordinate (d) at ($ (c) -2*#3*\ny*(1,0) $);
				\node at ($(d) +(0,#3*\nx)-0.2*(\nx/\absNx,0)$) {{\color{#5}#6}}; % 1 (d,b,c)
				\else
				\coordinate (d) at ($ (c) +2*#3*\nx*(0,1) $);
				\node at ($(d)-#3*\nx*(0,1)-0.2*\nx/\absNx*(1,0)$) {{\color{#5}#6}}; % 1
				\fi
				\draw[#5,line width=0.1mm] (d) -- (b); % horizontal line
				\draw[#5,line width=0.1mm] (d) -- (c); % vertical line
				\draw[#5,line width=0.1mm] (b) -- (c); % vertical line
} 
\newcommand{\coordSys}[9]{%(xo,yo,l,x1,y1,x2,y2,x3,y3)	
			\coordinate (center2) at (#1,#2); % center of figure
            \coordinate (x1) at ($ (center2) + #3*(#4,#5) $); % length=0.4
            \coordinate (x2) at ($ (center2) - .3*#3*(#4,#5) $); % length=0.4
            \coordinate (y1) at ($ (center2) + #3*(#6,#7) $);
            \coordinate (y2) at ($ (center2) - .3*#3*(#6,#7) $);
            \coordinate (z1) at ($ (center2) + #3*(#8,#9) $); %arrow length
            \coordinate (z2) at ($ (center2) - .3*#3*(#8,#9) $); %arrow length
            \draw[->,>=stealth,black][line width=0.4mm] (x2) -- (x1);
			\node at ($(x1)+(0.2,0)$){{\color{black} \large $x$}};
            \draw[->,>=stealth,black][line width=0.4mm] (y2) -- (y1);
            \node at ($(y1)+(0,0.2)$){{\color{black} \large $y$}};
            \draw[->,>=stealth,black][line width=0.4mm] (z2) -- (z1);
			\node at ($(z1)+(0,0.2)$){{\color{black} \large $z$}};
    	}
    	
\newcommand*{\Labelxy}[4]{\put(#1,#2) {\setlength{\fboxsep}{0pt}{\strut\textcolor{black}{\begin{turn}{#3}{#4}\end{turn}}}}}
\newcommand*{\LabelFig}[3]{\put(#1,#2) {\setlength{\fboxsep}{0pt}\colorbox{white}{\textcolor{black}{#3}}} }
\definecolor{darkolivegreen}{rgb}{0.33, 0.42, 0.18}
\definecolor{darkspringgreen}{rgb}{0.09, 0.45, 0.27}
\definecolor{darkslategray}{rgb}{0.18, 0.31, 0.31}
\definecolor{darkred}{rgb}{0.55, 0.0, 0.0}
\addeditor{kk}
\newcommand{\rhoico}{$\rho_\text{ico}$~} %{$\langle\rho^0_\text{ico}\rangle~$}
\newcommand{\dmin}{$D^2_\text{min}~$}
\newcommand{\glfive}{Co$_5$Cr$_2$Fe$_{40}$Mn$_{27}$Ni$_{26}$~}
\newcommand{\glfour}{CoNiCrFeMn~}

\newcommand{\gltwo}{CoNiCrFe~}
\newcommand{\glone}{CoNiFe~}

\newcommand{\gmax}{\gamma_{\text{max}}}
\newcommand{\hmin}{$h_{\text{min}}$~}
\newcommand{\tage}{$t_\text{age}$~}

\usepackage{xr}
\makeatletter
\newcommand{\cf}{{\it cf.~}}
\newcommand{\eg}{{\it eg.~}}

\newcommand*{\addFileDependency}[1]{
  \typeout{(#1)}
  \@addtofilelist{#1}
  \IfFileExists{#1}{}{\typeout{No file #1.}}
}
\makeatother
\newcommand*{\myexternaldocument}[1]{
    \externaldocument{#1}
    \addFileDependency{#1.tex}
    \addFileDependency{#1.aux}
}
\myexternaldocument{./sm}

\begin{document}

\title{Tuning brittleness in multi-component metallic glasses through chemical disorder aging}% Force line breaks with \\

\author{Kamran Karimi$^1$}
\email{kamran.karimi@ncbj.gov.pl}
\author{Stefanos Papanikolaou$^1$}
\email{stefanos.papanikolaou@ncbj.gov.pl }
\affiliation{%
 $^1$ NOMATEN Centre of Excellence, National Center for Nuclear Research, ul. A. Sołtana 7, 05-400 Swierk/Otwock, Poland\\
}%

\begin{abstract}
Shear localization in slowly-driven bulk metallic glasses (BMGs) is typically accompanied by a sharp drop in  the bulk stress response as a signature of the plastic yielding transition.
It is also observed that the sharpness of this elastic-plastic dynamical transition  depends on the extent of local chemical and microstructural orders, as well as the glass preparation protocol ({
\it ie}. thermal annealing).
Here, we investigate sheared multi-element BMGs in molecular dynamics (MD) simulations, and  demonstrate that  glass aging, implemented through a hybrid Monte-Carlo(MC)-MD process,  sharpens the elastic-plastic transition through a distinct crossover, seen in strain patterns that gradually shift from diffuse features in as-quenched samples to localized (yet system-spanning) patterns in well-annealed glasses.
This effect of glass aging on the elastic-plastic transition is found to be  correlated to the inherent interplay between aging-induced icosahedra ordering and co-operative formation of shear transformation zones.
The observed crossover is quantified through a measure of the age-dependent susceptibility to plastic rearrangements, exhibiting strong (anti-)correlations to local ordering features, and the corresponding spatial correlation length grows with the aging timescale.
\end{abstract}

\maketitle
Plastic yielding in slowly sheared metallic glasses (below the glass transition temperature $T_g$) typically occurs via localization of intense, irrecoverable deformation, yet without crushing or crumbling within the bulk.
Microstructurally, this flow is  attributed to  emergent shear transformation zones (STZs) \cite{argon1979plastic,falk1998dynamics, falk2011deformation, karimi2018correlation}, commonly known as carriers of amorphous plasticity, somewhat analogously to  dislocations in crystals.
STZs mutually interact, as they are soft mesoscale defects that relax stress locally but can further induce far-field elastic-type triggering elsewhere within the glassy medium.
This phenomenon leads to \emph{collective} dynamics upon failure and \emph{universal} features, including scale-free statistics and diverging length, time, and/or energy scales that could be understood within the broad context of far-from-equilibrium critical phenomena \cite{ozawa2018random,denisov2017universal,karimi2017inertia,karimi2019plastic}.
Despite the observed universality, the  \emph{sharpness} of the elastic-plastic transition (seen for example, as a discontinuous stress drop in displacement-controlled uniaxial loading
), may exhibit significant variations across multi-element BMGs owing to modifications in thermal treatments (i.e. aging/annealing) and chemical compositions \cite{cheng2011atomic,cheng2008local,cheng2009correlation,kim2018role}. 
In this paper, we concentrated on the equiatomic \glfour alloy that has been commonly regarded in the literature as a high-entropy ``Cantor" alloy~\cite{li2018microstructures,li2019mechanical} but we consider only its mechanical properties in the glassy state achieved by fast cooling~\cite{lu2017cooling}. 
We investigate the inherent correlations between microstructure and shear localization, as the amorphous state is gradually aging.

Amorphous metals have the capability to undergo plastic flow mediated by STZs, contributing significantly to their ductility \cite{wang2018spatial, wu2009transition, chen2006extraordinary, karimi2021shear}. 
However, in certain aged glasses, which do not possess  inherent heterogeneities \cite{cheng2008local, shi2005strain, albano2005shear}, or when the associated length scales do not significantly surpass the average interatomic distance, plastic distortion  displays localization within a dominant band and eventually  brittle-type fracture.
This is akin to the ductile-to-brittle transition present in a broad range of amorphous solids \cite{ozawa2018random}.
 In addition, aging-mediated, structural relaxation, leads to ``annealed" metallic glasses that nucleate certain quasi-ordered phases characterized by short range order (SRO)  \cite{ding2014full,ding2012correlating,ding2014soft,ma2016tailoring}, 
 and there exists a strong tendency to tune the extent of shear localization through structural ordering \cite{shi2005strain,shi2006atomic,shi2007stress}.
SROs are commonly identified by the formation of ordered icosahedral clusters, representing the most (energetically) favored atomic arrangement within the chemically diverse, amorphous matrix. 
Such clusters play a pivotal role in developing a number of glassy properties, such as the dynamic slowing-down in the super-cooled regime~\cite{hufnagel2016deformation}.
It is worth noting that SROs serve as ``infertile" sites for the nucleation of STZs in that the latter are typically loosely-packed, soft, and disordered arrangements that weaken the local strength and, thus, enhance the nucleation probability of shear banding instabilities.

In this paper, we investigate the precise characteristics of the inherent SRO-STZ interplay and relevant atomistic mechanisms that influence the nucleation dynamics of shear bands in concentrated and chemically complex BMGs. 
Previously \cite{karimi2021shear,karimi2023yielding}, in concentrated BMGs, we showed how the elastic-plastic transition sharpness and fine-scale structural ordering features in shear bands might be dependent on the chemical composition.
Here, we show how room temperature thermal aging may affect four chemically complex glasses and then, we focus on the structural aspects of the transition in the   \glfour metallic glass.
We center our approach on a classification pertinent to the plastic yielding transition in metallic glasses: \emph{i}) the  ``good" or annealed glass, forms localized deformation patterns and displays a discontinuous transition in the macroscopic average stress response, \emph{ii}) the ``bad" or quenched glass, may delocalize strain and display ductility characteristics.
We demonstrate that annealing  effectively controls the crossover from bad to good glass by tuning the level of icosahedra-based structural ordering within the glass, thereby influencing its propensity to form shear bands.
We perform cluster analysis for a measure that measures non-affine plastic rearrangements, and we show that the cluster correlation length evolves across the yielding transition and exhibits significant variations with sample age.\\

%%%%%%%%%%%%%%%%%%%%%%%%%%%%%%%%%%%%%%%%%%%%%%%%%%%%%%%%%%%%%%%%%%%%
%%%%%%%%%%% figure
%%%%%%%%%%%%%%%%%%%%%%%%%%%%%%%%%%%%%%%%%%%%%%%%%%%%%%%%%%%%%%%%%%%%        
\begin{figure}[t]
    \centering
    \begin{overpic}[width=0.49\columnwidth]{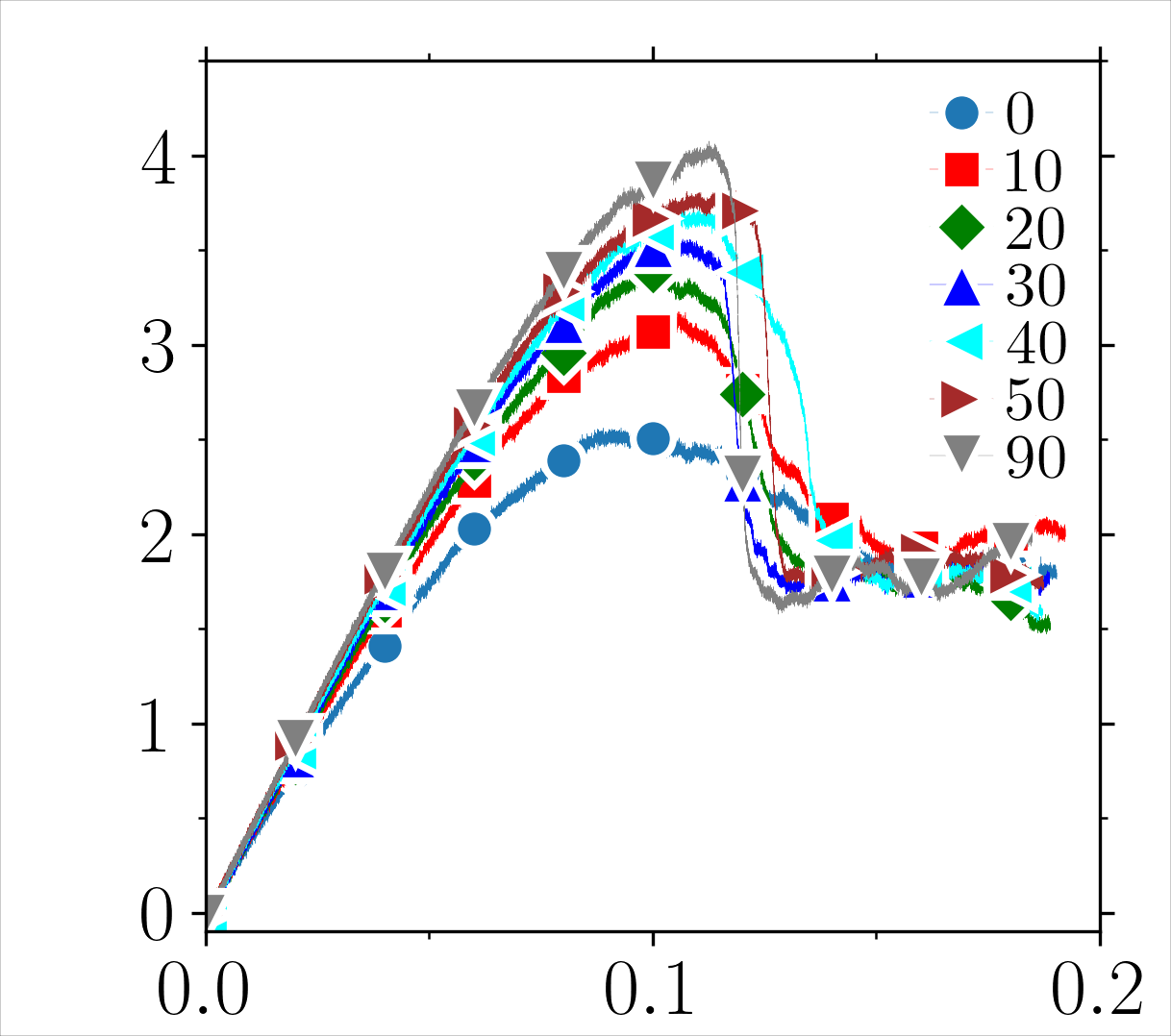}
        \LabelFig{19}{87}{$a)$ \scriptsize \glfour }
        \Labelxy{50}{-5}{0}{$\gamma_{xy}$}
        \Labelxy{3}{35}{90}{{$\sigma_{xy}$ \tiny(Gpa)}}
        \Labelxy{72}{49}{90}{ $\scriptstyle t_\text{age}(\text{ps})$ }
      \end{overpic}
    \begin{overpic}[width=0.49\columnwidth]{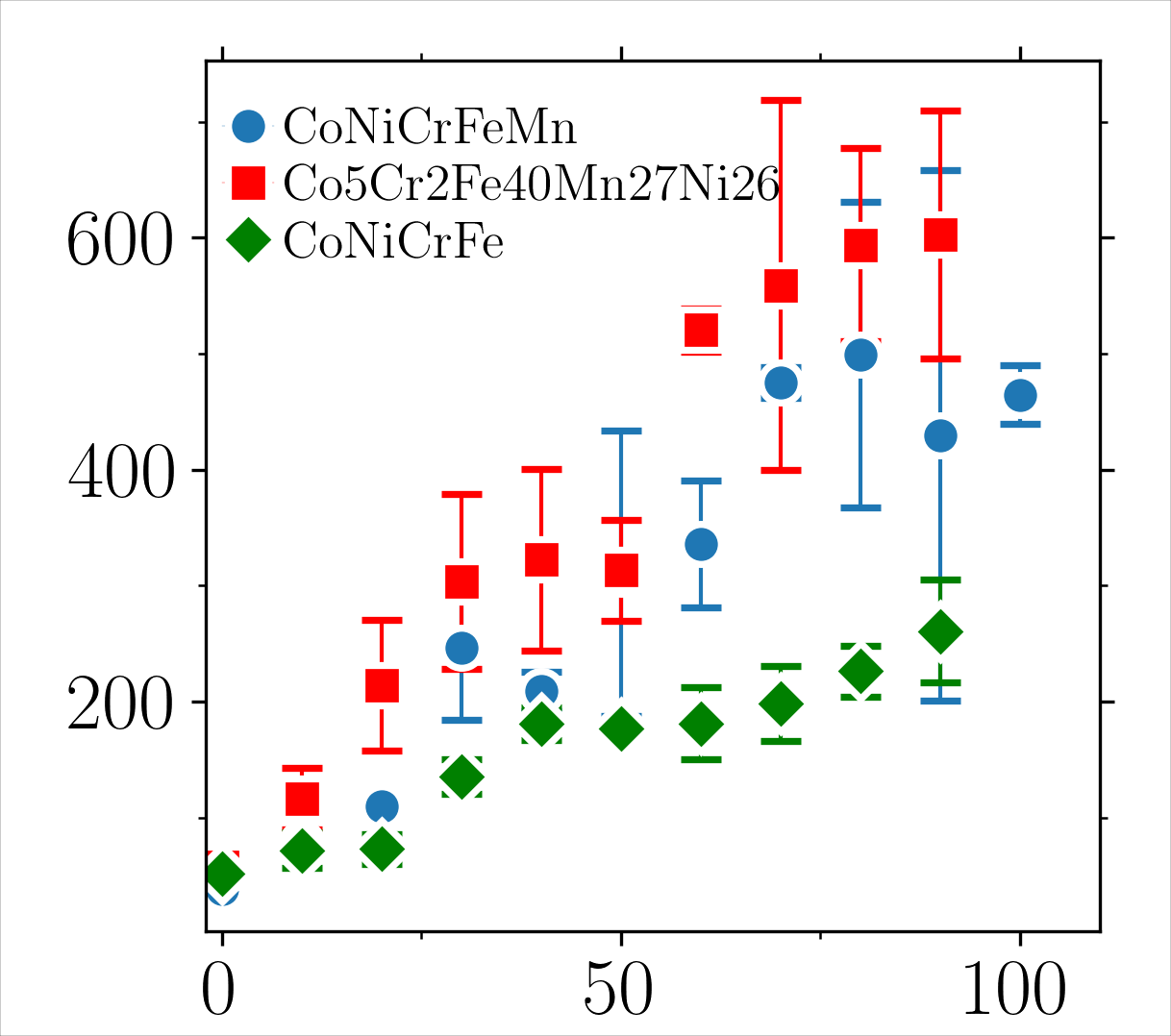}
        \LabelFig{19}{87}{$b)$}
        \Labelxy{50}{-5}{0}{$t_\text{age}$ (ps)}
        \Labelxy{-3}{35}{90}{{\hmin\tiny(Gpa)}}
     \end{overpic}
    \vspace{12pt}

    \begin{overpic}[width=0.23\textwidth]{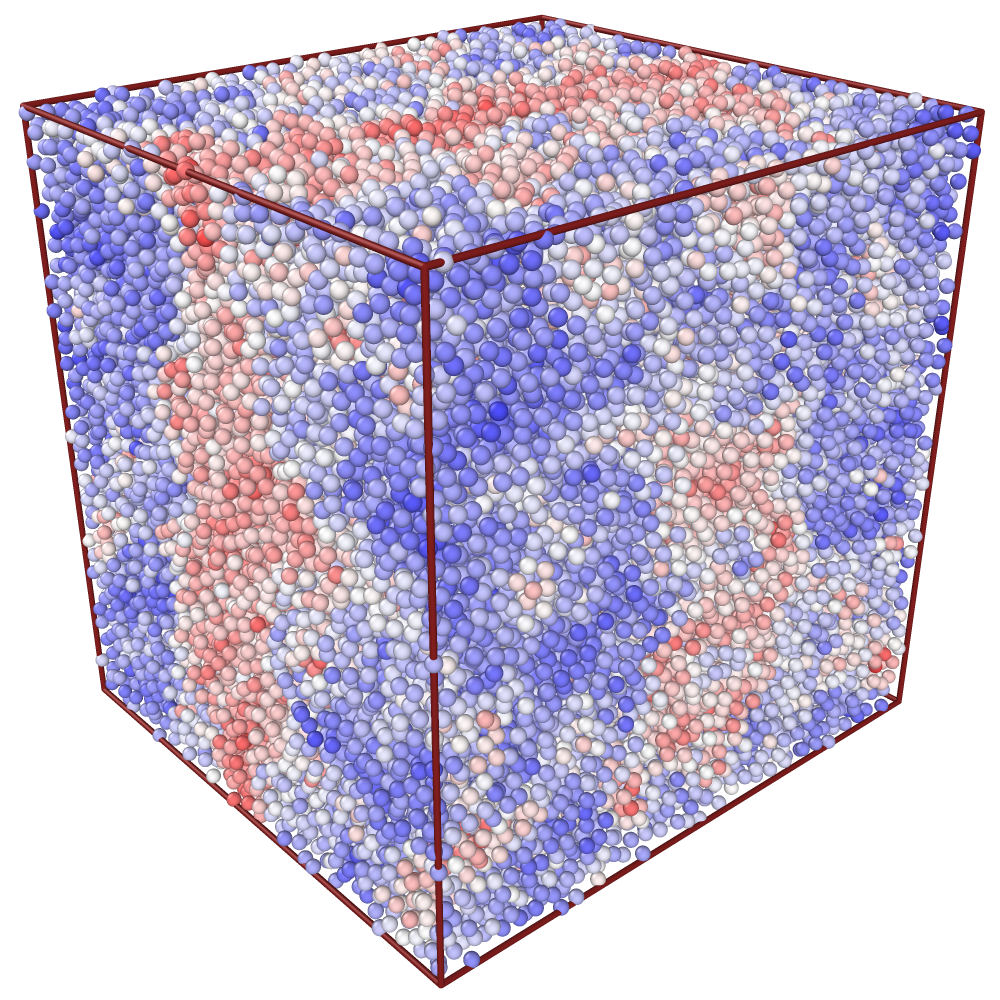}
        \LabelFig{18}{102}{$c)$ \scriptsize As-quenched}
        % color bar
        \put(80,10){\includegraphics[height=0.15cm,width=1.7cm]{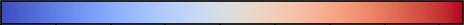}}  
        \put(80,3) {\footnotesize Low~~~~~~~High}
        \put(94,19) {\footnotesize \dmin}
        % coord. sys.
       \begin{tikzpicture}
            \coordinate (a) at (0,0); % reference
             \node[white] at (a) {\tiny.};
             \coordSys{0}{0}{.55}{1}{-0.6}{1}{1}{-0.2}{1} 
        \end{tikzpicture}	 

    \end{overpic}
    \begin{overpic}[width=0.23\textwidth]{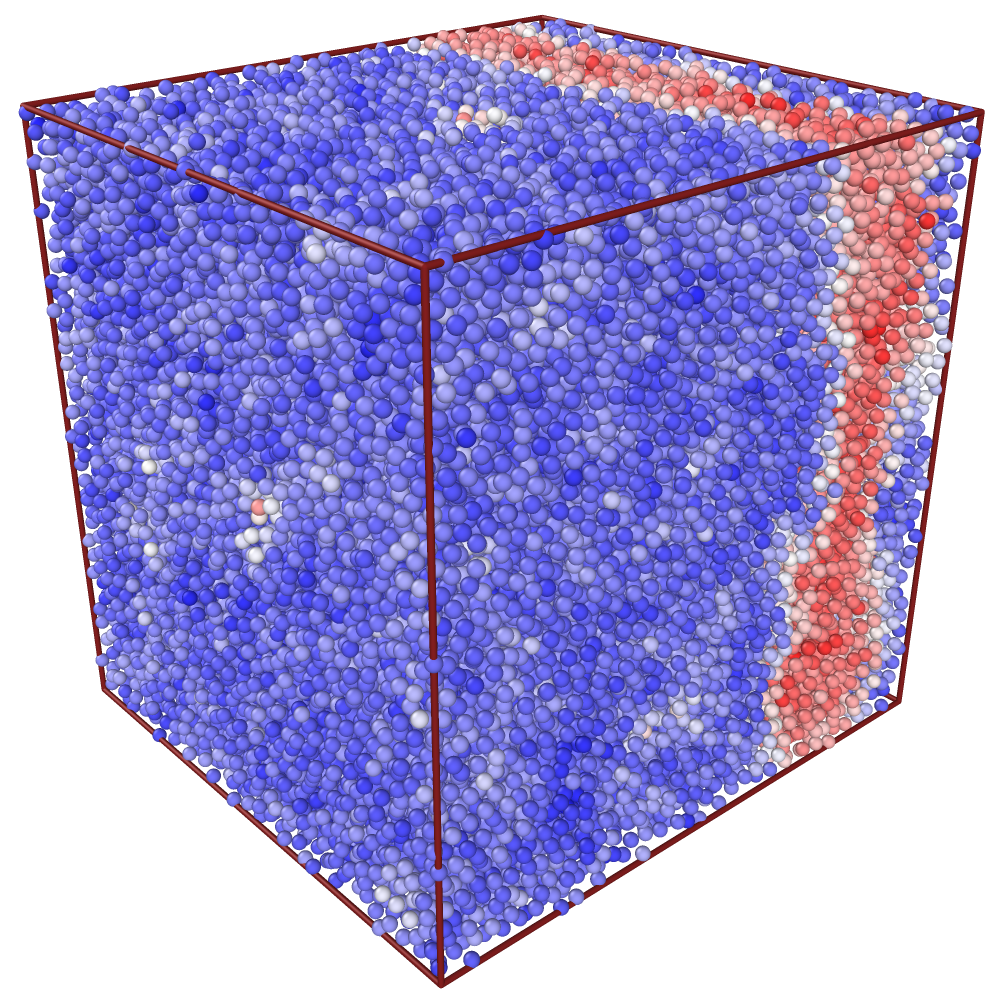}
        \LabelFig{18}{102}{$d)$ \scriptsize Annealed}
    \end{overpic}
    \caption{\textbf{a}) Macroscopic stress $\sigma_{xy}$ plotted against applied (shear) strain $\gamma_{xy}$ in deforming \glfour annealed at different duration $t_\text{age}$ \textbf{b}) softening modulus \hmin versus annealing duration $t_\text{age}$ corresponding to several BMGs. Local mean-squared nonaffine displacements \dmin associated with \textbf{c}) as-quenched ($t_\text{age}=0$) and \textbf{d}) annealed ($t_\text{age}=90$ ps) \glfour at $\gamma_{xy}=0.2$. Here $x$, $y$, and $z$ denote flow, gradient, and vorticity directions, respectively. The blue and red colors in \textbf{c}) and \textbf{d}) indicate low and high \dmin values, respectively. The size of the cubic box is $L\simeq 80$ \r{A}.}
    \label{fig:loadCurve}
\end{figure}
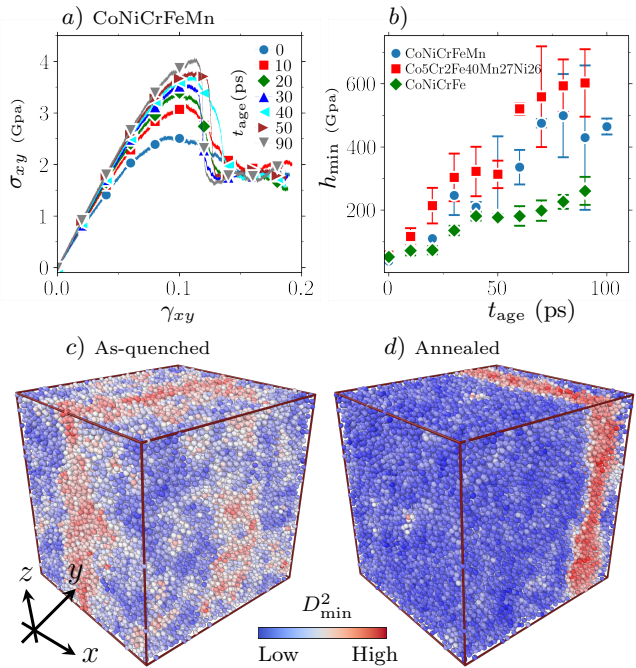

\noindent\emph{Simulations \& Protocols~--~}The details of the  hybrid Monte Carlo (MC) - Molecular Dynamics (MD) simulations, in this work, are given in the Supplementary Material (SM) \footnote{See Supplemental Material at [URL will be inserted by publisher] for further discussions relevant to simulation details.} (see also references \cite{plimpton1995fast,ding2013high,alvarez2023viewing,choi2018understanding,lees1972computer} therein) 
including the description of relevant units, the preparation protocol, the utilized interatomic  potential functions, and also, the deformation parameters of the investigated model metallic glasses \cite{karimi2021shear}.
Prior to shearing, the as-quenched samples were subject to MC-MD annealing up to the aging duration of \tage$=100$ ps (with standard heuristic assumptions on the MC-related timescale).
Simple shear tests were subsequently performed on the aged glasses at a fixed strain rate $\dot{\gamma}_{xy}=10^{-4}~\text{ps}^{-1}$ and temperature $T=300$ K,  up to shear strain $\gamma_{xy} = 0.2$.  
To probe the dynamics of individual atoms, we track the mean squared displacements \dmin as a measure of atoms' non-affinity with respect to the imposed shear deformation \cite{falk1998dynamics}.
We further perform a Voronoi analysis using OVITO \cite{stukowski2009visualization} to locate atoms displaying  icosahedral order, namely polyhedral cells with exactly twelve faces, including regular pentagons.
To obtain the associated (number) density, \rhoico$=1/V_\text{ico}$, we repeat the Voronoi analysis by including \emph{exclusively} atoms with icosahedral symmetries within the periodic box (and excluding  other atoms).
This gives another set of Voronoi cells with volume $V_\text{ico}$.
The mean number density of icosahedral clusters is further derived as $\langle$\rhoico\!$\rangle$.

%This methodology allows us to probe the spatial-dynamical evolution of the local shear modulus $\mu=c_{xyxy}$ and associated percolation features near failure transition in sheared glasses. 
Further, we measure the softening modulus \hmin\!\!\!\!, defined as the maximum \emph{rate} of the macroscopic average stress drop at every age \tage as in Fig.~S3. %\ref{fig:hminIllustration}.
This stress drop is typically defined as the difference between the overshoot stress and the subsequent flow stress, and is associated with the initiation of a spanning shear band, thus it has been used as an order parameter in model glass studies \cite{ozawa2018random,chen2011theory} showing meaningful variations with glass compositions and processing parameters \cite{cheng2008local}.
Nevertheless, in metallic glass simulations and/or experiments, a robust measurement of the macroscopic drop is not always feasible due to the lack of a well-defined steady flow regime beyond the apparent stress overshoot.
In a recent work \cite{karimi2021shear}, we established \hmin as a more robust experimentally-relevant indicator of the elastic-plastic transition in BMGs, shear banding and associated structural features.\\

%%%%%%%%%%%%%%%%%%%%%%%%%%%%%%%%%%%%%%%%%%%%%%%%%%%%%%%%%%%%%%%%%%%%
%%%%%%%%%%% figure
%%%%%%%%%%%%%%%%%%%%%%%%%%%%%%%%%%%%%%%%%%%%%%%%%%%%%%%%%%%%%%%%%%%%
\begin{figure}[t]
    \begin{overpic}[width=0.49\columnwidth]{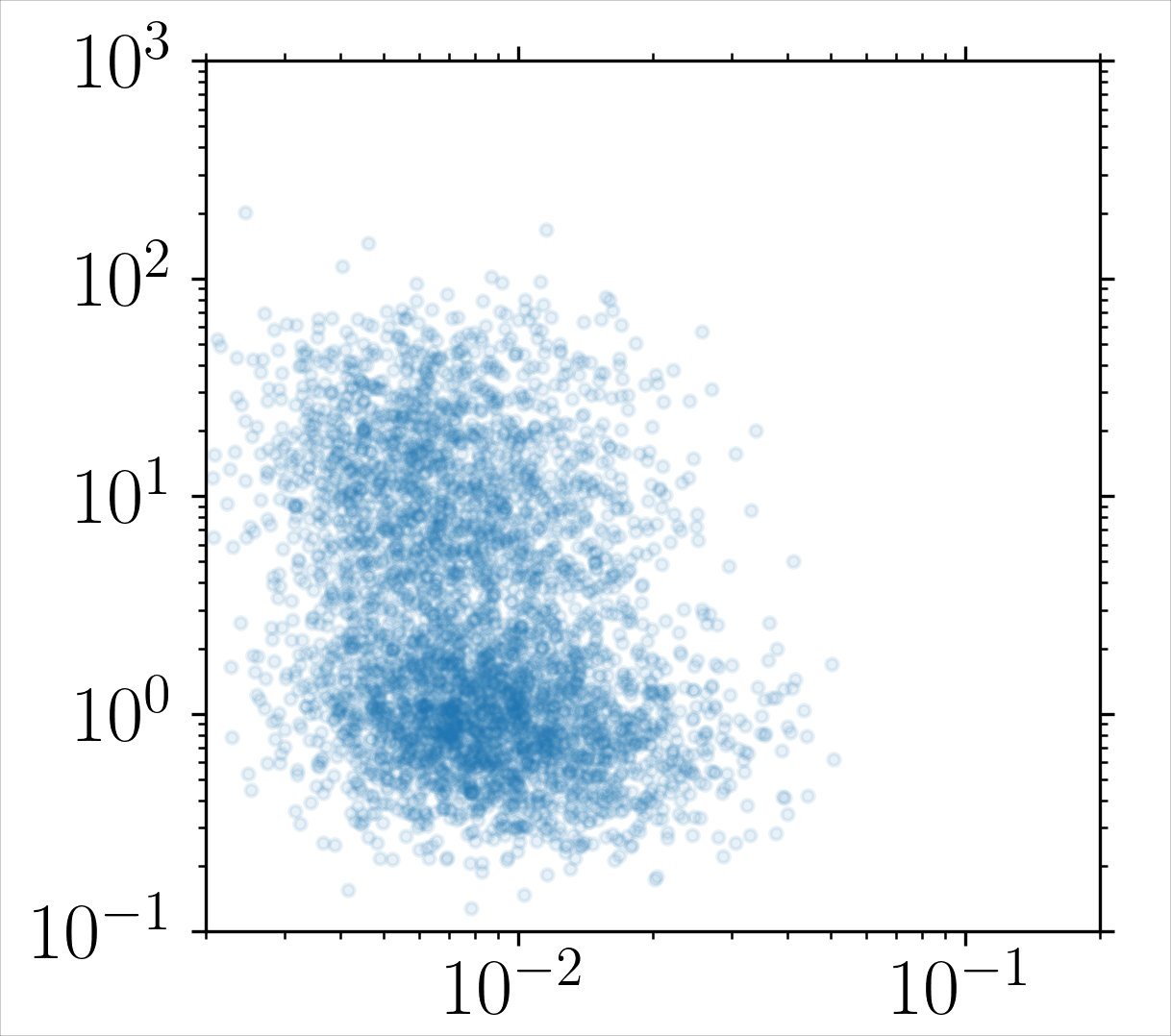}
        \Labelxy{-4}{35}{90}{\dmin (\r{A}$^2$)}
        \Labelxy{38}{-7}{0}{$\rho_\text{ico}$ (\r{A}$^{-3}$)}
         \LabelFig{19}{76}{$a)$~\scriptsize As-quenched}
         \put(66,50){\includegraphics[width=0.2\columnwidth]{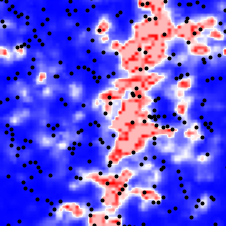}}
    \end{overpic}
     \begin{overpic}[width=0.49\columnwidth]{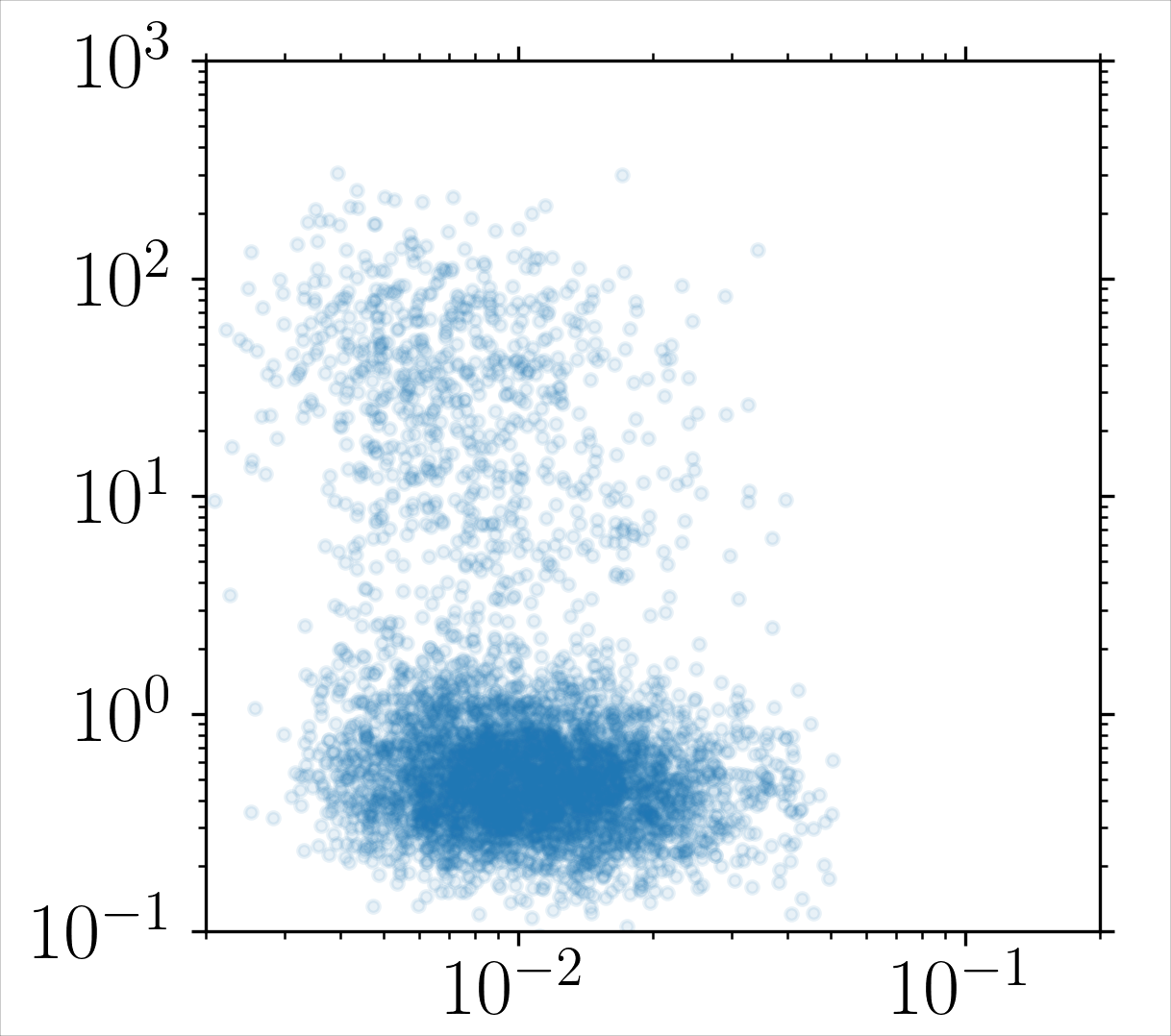}
        \Labelxy{38}{-7}{0}{$\rho_\text{ico}$ (\r{A}$^{-3}$)}
        \put(62,50){\includegraphics[width=0.2\columnwidth]{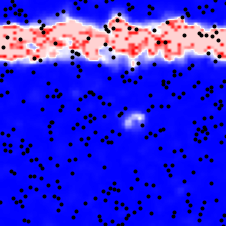}}
        \LabelFig{19}{76}{$b)$~\scriptsize Annealed}
    \end{overpic}
    \vspace{+2pt}

    \begin{overpic}[width=0.49\columnwidth]{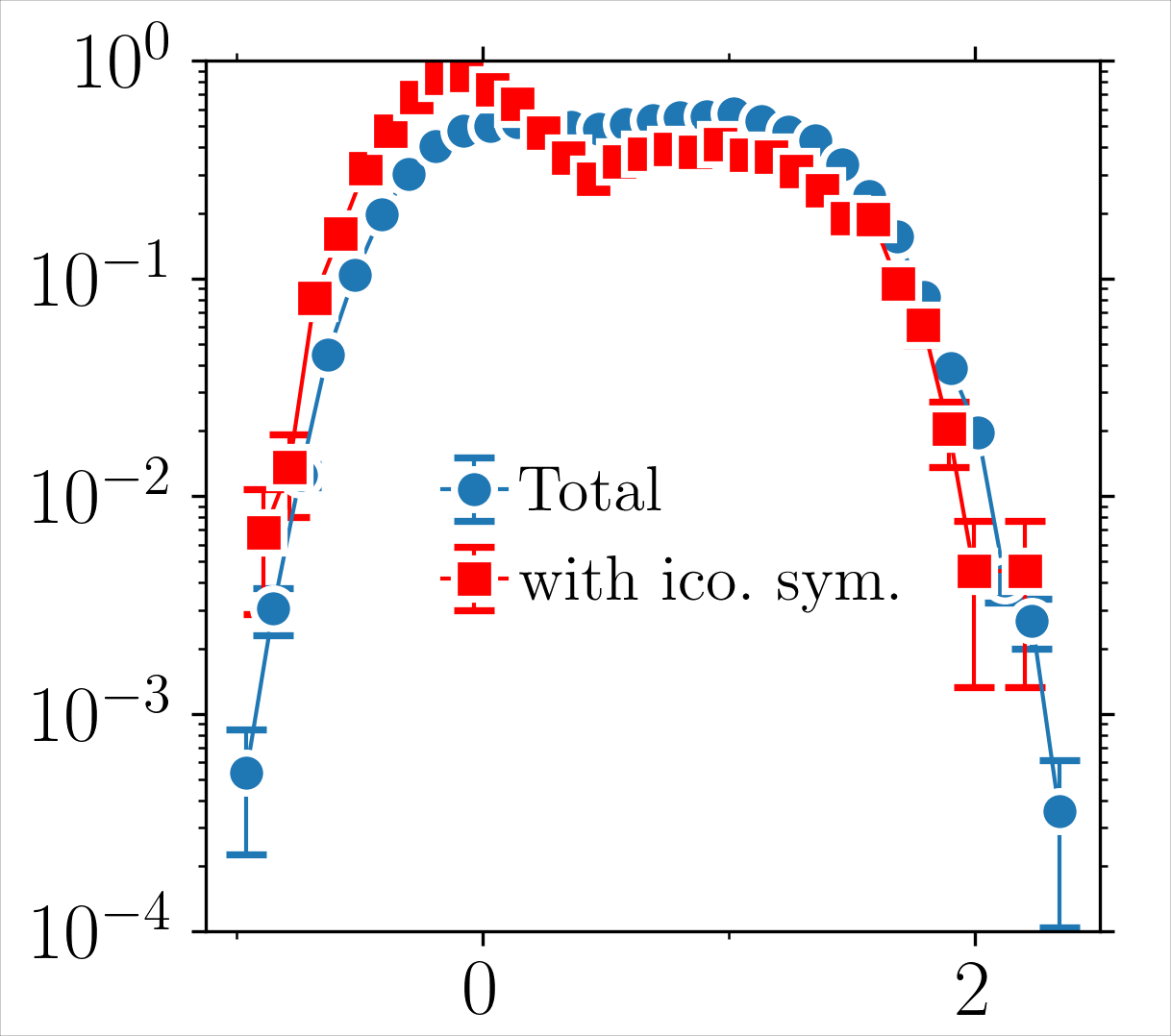}
        \Labelxy{-6}{30}{90}{Probability} %$P(\text{Log}_{10}$\dmin\!)}
        \Labelxy{38}{-6}{0}{$\text{Log}_{10}$\dmin}
         \LabelFig{19}{13}{$c)$ \scriptsize As-quenched}
    \end{overpic}
    \begin{overpic}[width=0.49\columnwidth]{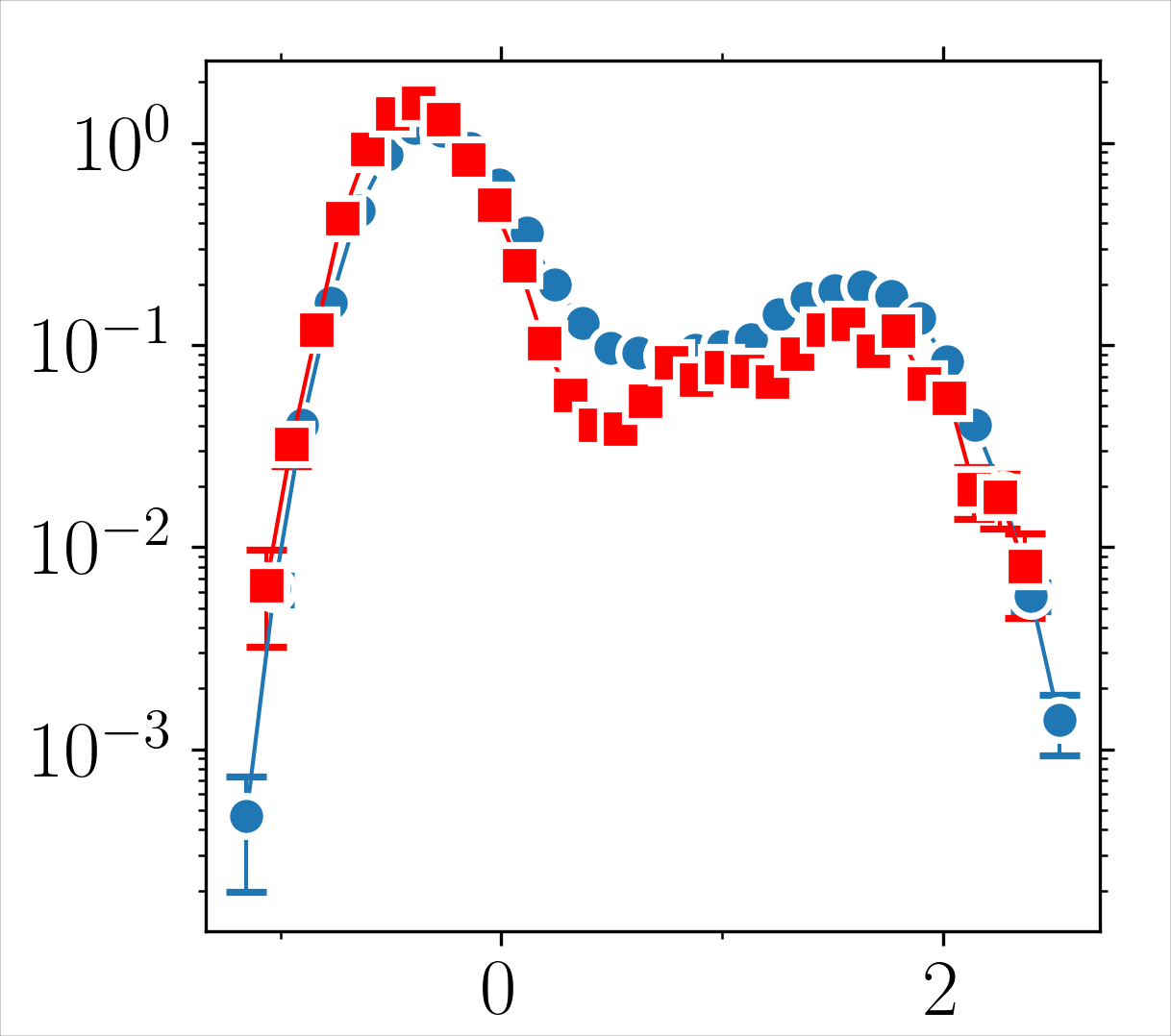}
        \Labelxy{-6}{30}{90}{Probability}% $P(\text{Log}_{10}$\dmin\!)}
        \Labelxy{38}{-6}{0}{$\text{Log}_{10}$\dmin}
        \LabelFig{19}{13}{$d)$ \scriptsize Annealed}
    \end{overpic}
    \caption{Scatter plots of $D^2_\text{min}$ and icosahedra density $\rho_\text{ico}$ in the \textbf{a}) as-quenched ($t_\text{age}=0$) and \textbf{b}) annealed ($t_\text{age}=90$ ps) \glfour at $\gamma_{xy}=0.2$. $D^2_\text{min}$ probability distribution function of atoms with the full icosahedral order corresponding to the \textbf{c}) as-quenched and \textbf{d}) annealed glasses at $\gamma_{xy}=0.2$. The corresponding $D^2_\text{min}$ spatial maps are shown in the insets with the blue and red colors indicating spatial regions with low and high nonaffine displacements. The (black) dots denote atoms with full icosahedral order. The scale of the color map is $L\simeq 80$ \r{A}.}
    \label{fig:scatterD2minRho}
\end{figure}

\noindent\emph{Results~--~}Figure~\ref{fig:loadCurve} displays results of the shear tests performed on the aged \glfour glass.
The resulting stress-strain curves, $\sigma_{xy}$ against $\gamma_{xy}$ in Fig.~\ref{fig:loadCurve}(a), indicate a pronounced stress overshoot, namely a monotonic increase of stress towards a peak value at $\gmax \simeq 0.1$,  followed by a sheer reduction in stress, and prior to a well-defined plastic flow regime, with  marked dependence on glass aging.
This is quantified in Fig.~\ref{fig:loadCurve}(b) where the rate of stress drop \hmin tends to grow (and eventually saturate) with increasing $t_\text{age}$ corresponding to \glfour as well as \glfive\!\!, \gltwo\!\!, and \glone\!\!.
This feature appears to be quite robust with respect to variations in the chemical composition and/or molar concentrations.
The observed enhancement in the sharpness of yielding transition aligns closely with the \dmin maps as visualized in Fig.~\ref{fig:loadCurve}(b) and (c) corresponding to the as-quenched ($t_\text{age}=0$) and annealed ($t_\text{age}=90$ ps) glasses at $\gamma_{xy}=0.2$.
Notably, the abrupt stress drop in the aged metallic glass is accompanied by localized (and system-spanning) features as illustrated in Fig.~\ref{fig:loadCurve}(d), whereas the as-quenched sample displays scattered deformation patterns across the medium, as in Fig.~\ref{fig:loadCurve}(c).
Here, the blue and red colors in the deformation maps indicate regions with low and high squared nonaffine displacements.

We now turn to the characteristics of  microstructural ordering and the interplay with shear bands in the specific, for clarity purposes, deforming metallic glass \glfour. 
The color maps in the insets of Fig.~\ref{fig:scatterD2minRho}(a) and (b) overlay atoms with the full icosahedral order (black disks) on the two-dimensional (interpolated) \dmin field associated with the as-quenched (\tage$=0$ ps) and annealed (\tage$=90$ ps) metallic glass at $\gamma_{xy}=0.2$.
It is evident from both maps that (red) rearranging zones notably lack local structural ordering in contrast to the (blue) rigid matrix.
This is further illustrated in Fig.~\ref{fig:scatterD2minRho}(c) and (d) displaying \dmin probability distribution functions corresponding to atoms with icosahedral symmetry in the as-quenched and annealed sample, respectively.
The latter exhibits a clear bimodal behavior in Fig.~\ref{fig:scatterD2minRho}(d) with the first (higher) and second (lower) peaks denoting the population of atoms outside and within shear zones.
The conditional distribution (red squares) indicates a relatively higher contribution of the ordered icosahedral phase to the higher peak in very close agreement with the observation of rare ordering occurrences within plastically deforming zones.
Such features are also present in Fig.~\ref{fig:scatterD2minRho}(c) corresponding to the as-quenched glass but with less pronounced bimodality.

\begin{figure}[b]
%    \raggedright
     \begin{overpic}[width=0.49\columnwidth]{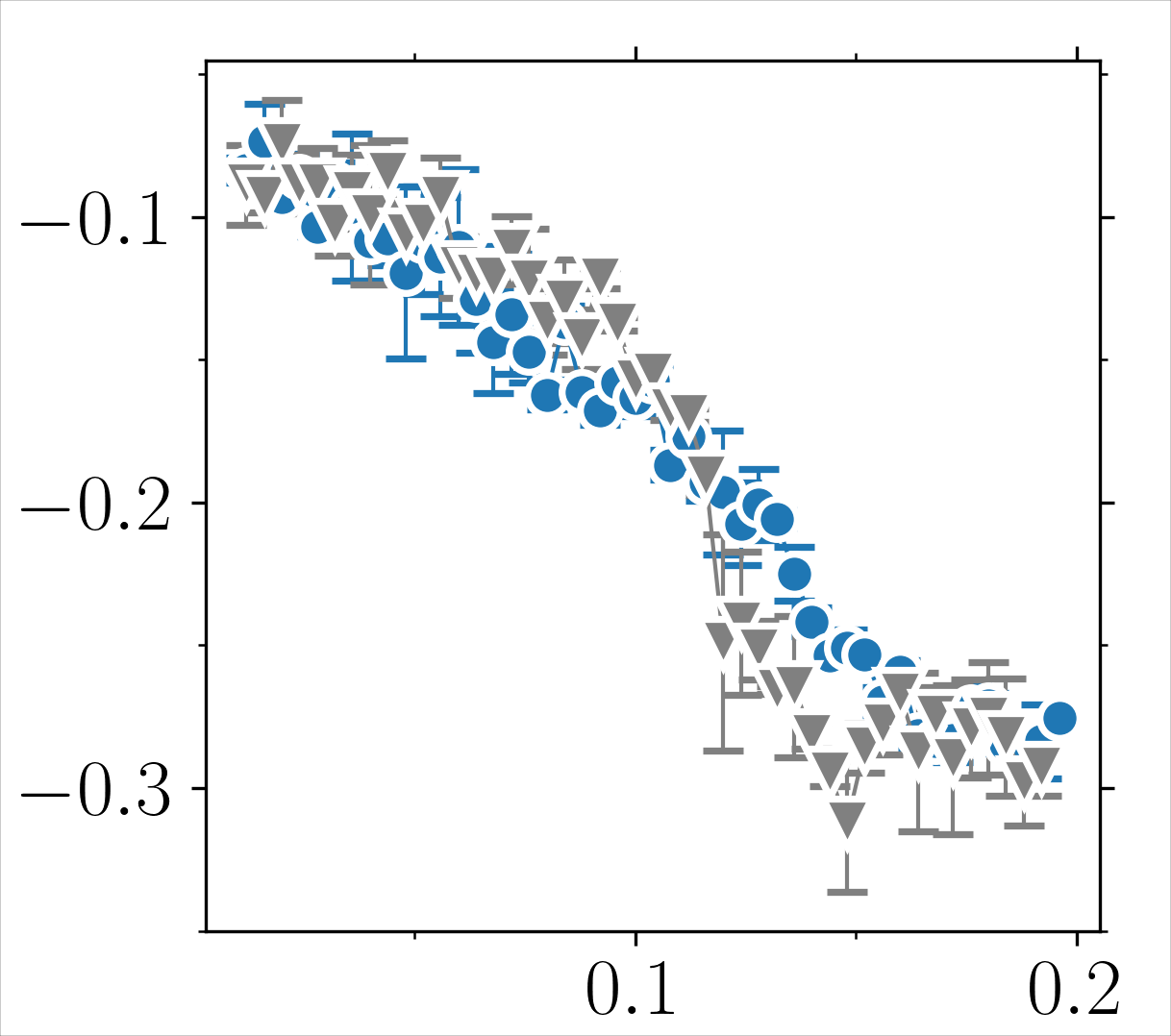}
        \Labelxy{50}{-3}{0}{$\gamma_{xy}$}
        \Labelxy{-4}{35}{90}{$c_{XY}$}
        % \Labelxy{72}{56}{90}{ $\scriptstyle t_\text{age}(\text{ps})$ }
        \LabelFig{19}{76}{$a)$}
    \end{overpic}
    \begin{overpic}[width=0.49\columnwidth]{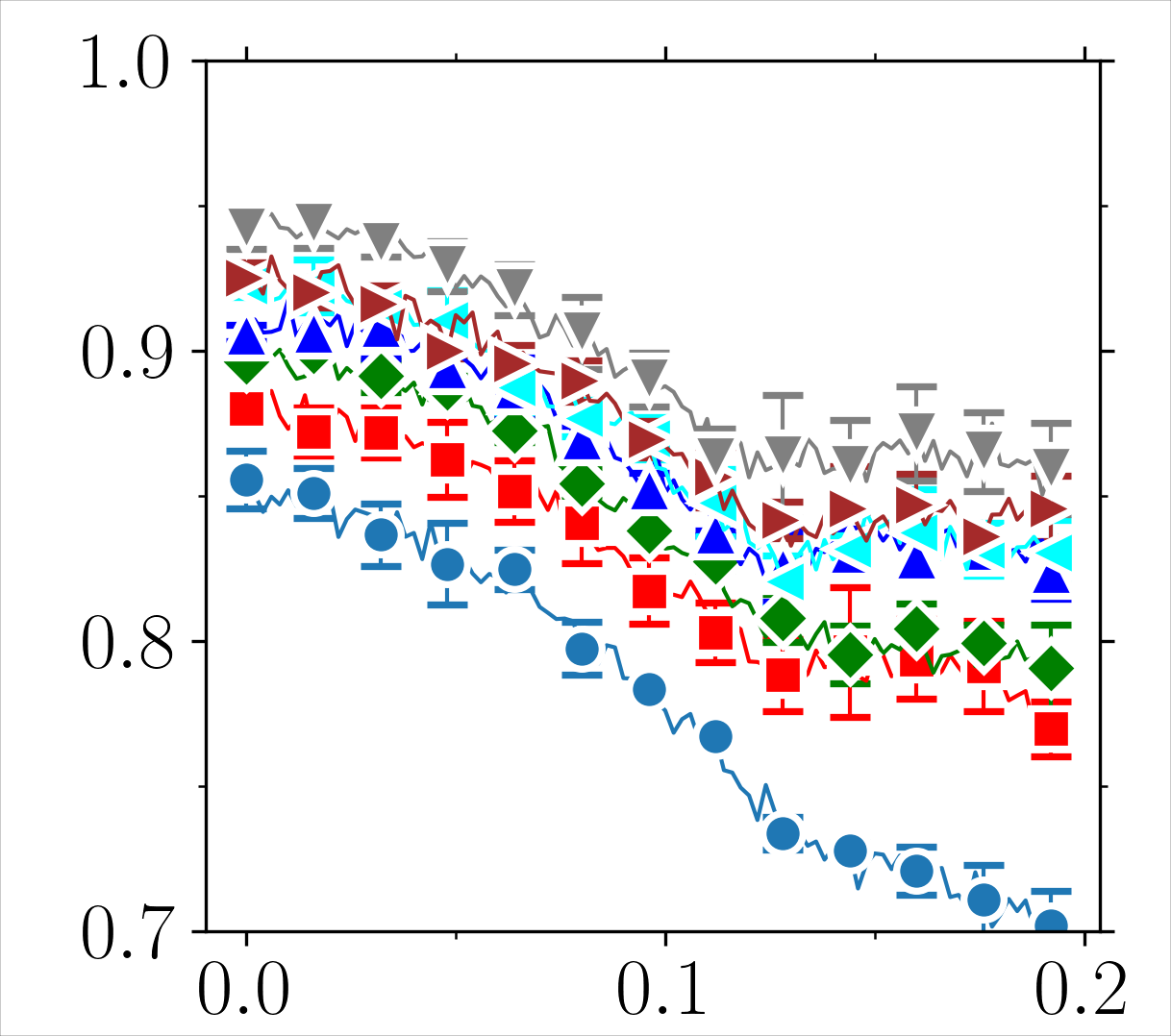}
        \LabelFig{19}{76}{$b)$}
        \Labelxy{50}{-3}{0}{$\gamma_{xy}$}
%        \Labelxy{-2}{32}{90}{$\langle$ \dmin $\rangle$ \tiny(\r{A}$^2$)}
        \Labelxy{-5}{32}{90}{$\langle$\rhoico\!$\rangle$ \tiny(\r{A}$^{-3}$)}
        \put(100,30){\includegraphics[width=0.07\columnwidth]{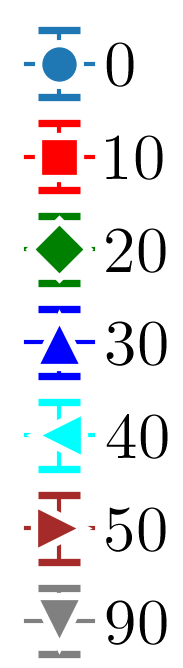}}  
        \Labelxy{100}{83}{0}{ $\scriptstyle t_\text{age}(\text{ps})$ }
    \end{overpic}
    \vspace{-4pt}
    
    \begin{overpic}[width=0.49\columnwidth]{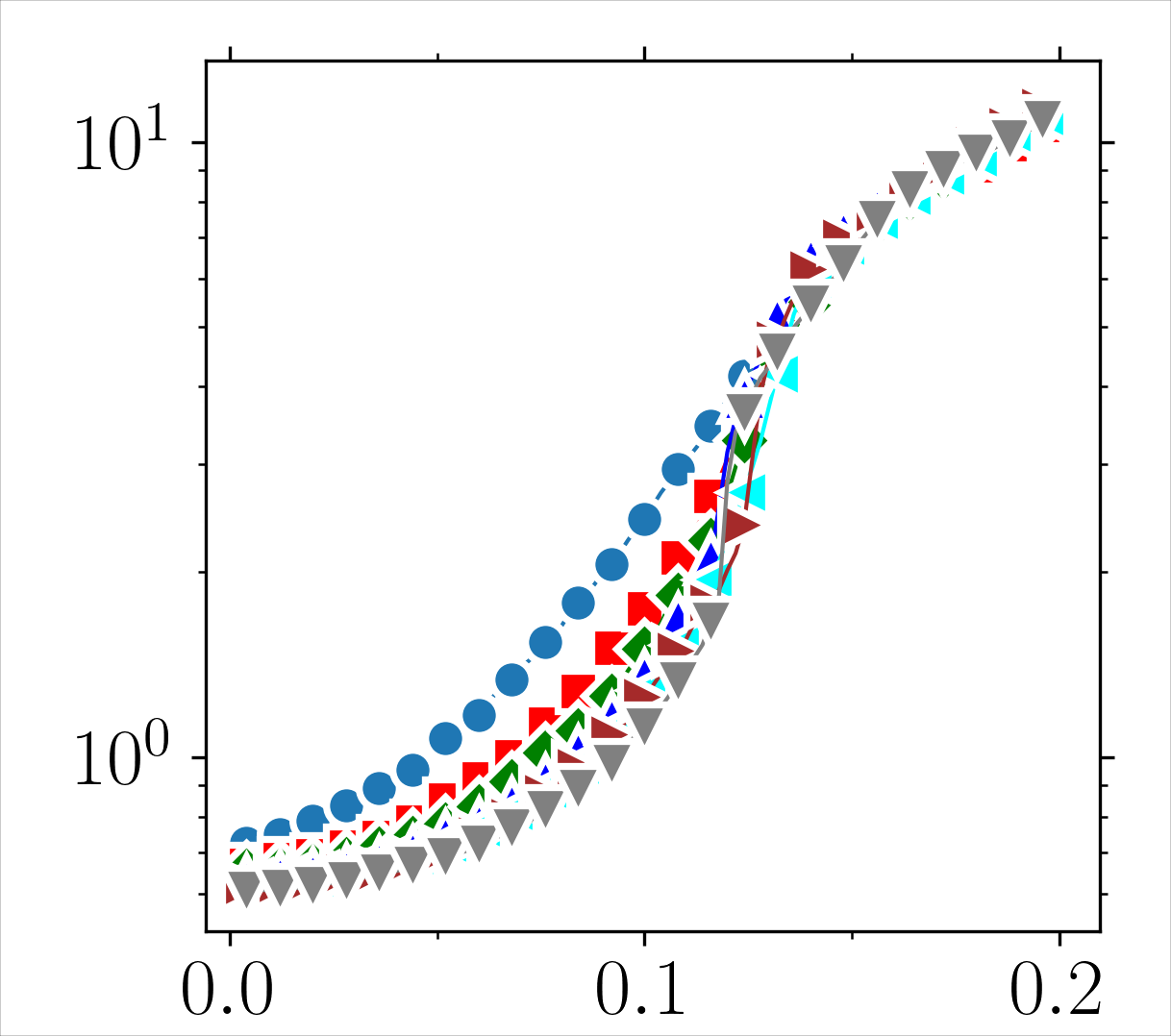}
        \Labelxy{50}{-4}{0}{$\gamma_{xy}$}
        \Labelxy{-2}{32}{90}{$\langle$\dmin\!\!$\rangle$ \tiny(\r{A}$^2$)}
        % \Labelxy{69}{16}{90}{ $\scriptstyle t_\text{age}(\text{ps})$ }
         \LabelFig{19}{76}{$c)$}
    \end{overpic}
    \begin{overpic}[width=0.49\columnwidth]{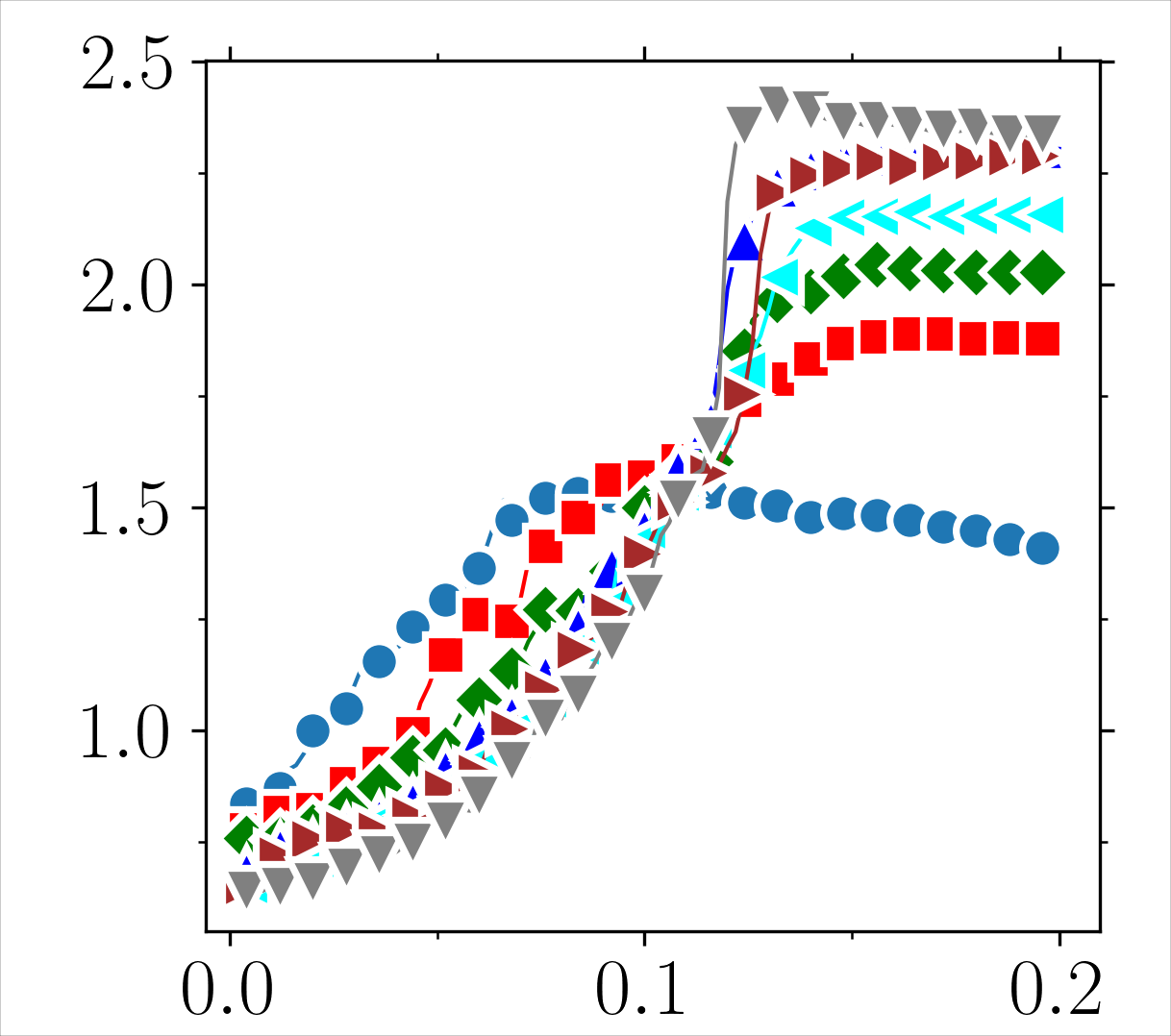}
        \Labelxy{50}{-4}{0}{$\gamma_{xy}$}
        \Labelxy{-5}{20}{90}{std(\dmin\!)/$\langle$\dmin\!\!$\rangle$}
         \LabelFig{19}{76}{$d )$}
    \end{overpic}
    \caption{Evolution of \textbf{a}) correlation coefficient $c_{XY}$ \textbf{b}) mean number density of atoms with icosahedral symmetries $\langle\rho^\text{ico}\rangle$ \textbf{c}) mean squared nonaffine displacements $\langle$\dmin\!\!$\rangle$ \textbf{d}) scaled standard deviation std(\dmin\!)/$\langle$\dmin\!\!$\rangle$ associated with atoms' \dmin plotted against applied strain $\gamma_{xy}$ in \glfour corresponding to different aging duration $t_\text{age}$. The error bars denote standard errors.}
    \label{fig:orderParameters}
\end{figure}

Next, we consider the atoms with  icosahedral ordering and we carry out a cross correlation analysis between associated squared non-affine displacements and the number density \rhoico\!\!.
The scatter data of \dmin and \rhoico in Fig.~\ref{fig:scatterD2minRho}(a) and (b) indicate significant (anti-)correlations between the two observables $X=\text{log}_{10}{D^2_\text{min}}$ and $Y=\text{log}_{10}\rho_\text{ico}$. The (linear) correlation coefficient $c_{XY}=\langle \hat{X} \hat{Y} \rangle$ and its evolution with strain is shown in Fig.~\ref{fig:orderParameters}(a) where $\langle.\rangle_i$ denotes averaging over the atom index $i$ and $\hat{X}$ indicates the deviation from the mean $\langle X\rangle_i$, normalized by the standard deviation associated with each variable.
Overall, (anti-)correlation monotonically grows with loading, and it saturates at the onset of the plastic flow regime.
The aging process leads to a  sharp elastic-plastic transition on approach to failure (see Fig.~\ref{fig:loadCurve}(a) and (b), with a very infrequent occurrence of structural icosahedral ordering within the shear bands (regions with large $ D^2_\text{min}$) (\cf Fig.~\ref{fig:scatterD2minRho}(d)). 

%%%%%%%%%%%%%%%%%%%%%%%%%%%%%%%%%%%%%%%%%%%%%%%%%%%%%%%%%%%%%%%%%%%%
%%%%%%%%%%% figure
%%%%%%%%%%%%%%%%%%%%%%%%%%%%%%%%%%%%%%%%%%%%%%%%%%%%%%%%%%%%%%%%%%%%

% %%%%%%%%%%%%%%%%%%%%%%%%%%%%%%%%%%%%%%%%%%%%%%%%%%%%%%%%%%%%%%%%%%%%
% %%%%%%%%%%% figure
% %%%%%%%%%%%%%%%%%%%%%%%%%%%%%%%%%%%%%%%%%%%%%%%%%%%%%%%%%%%%%%%%%%%%
% \begin{figure}[t]
%     \begin{overpic}[width=0.5\columnwidth]{Figs2nd/rhoIco_strain_age.png}
%         \Labelxy{50}{-3}{0}{$\gamma_{xy}$}
% %        \Labelxy{-2}{32}{90}{$\langle$ \dmin $\rangle$ \tiny(\r{A}$^2$)}
%         \Labelxy{-5}{32}{90}{$\langle$\rhoico\!$\rangle$ \tiny(\r{A}$^{-3}$)}
%         \put(100,30){\includegraphics[width=0.07\columnwidth]{Figs2nd/rhoIco_strain_age_legend.png}}  
%         \Labelxy{100}{83}{0}{ $\scriptstyle t_\text{age}(\text{ps})$ }
%     \end{overpic}
%     \caption{Mean number density of atoms with icosahedral symmetries $\langle\rho^\text{ico}\rangle$ plotted against applied strain $\gamma_{xy}$ corresponding to different aging duration $t_\text{age}$. The error bars denote standard errors.}
%     \label{fig:rhoico}
% \end{figure}

%%%%%%%%%%%%%%%%%%%%%%%%%%%%%%%%%%%%%%%%%%%%%%%%%%%%%%%%%%%%%%%%%%%%
%%%%%%%%%%% figure
%%%%%%%%%%%%%%%%%%%%%%%%%%%%%%%%%%%%%%%%%%%%%%%%%%%%%%%%%%%%%%%%%%%%
\begin{figure*}[t]
    \begin{overpic}[width=0.48\columnwidth]{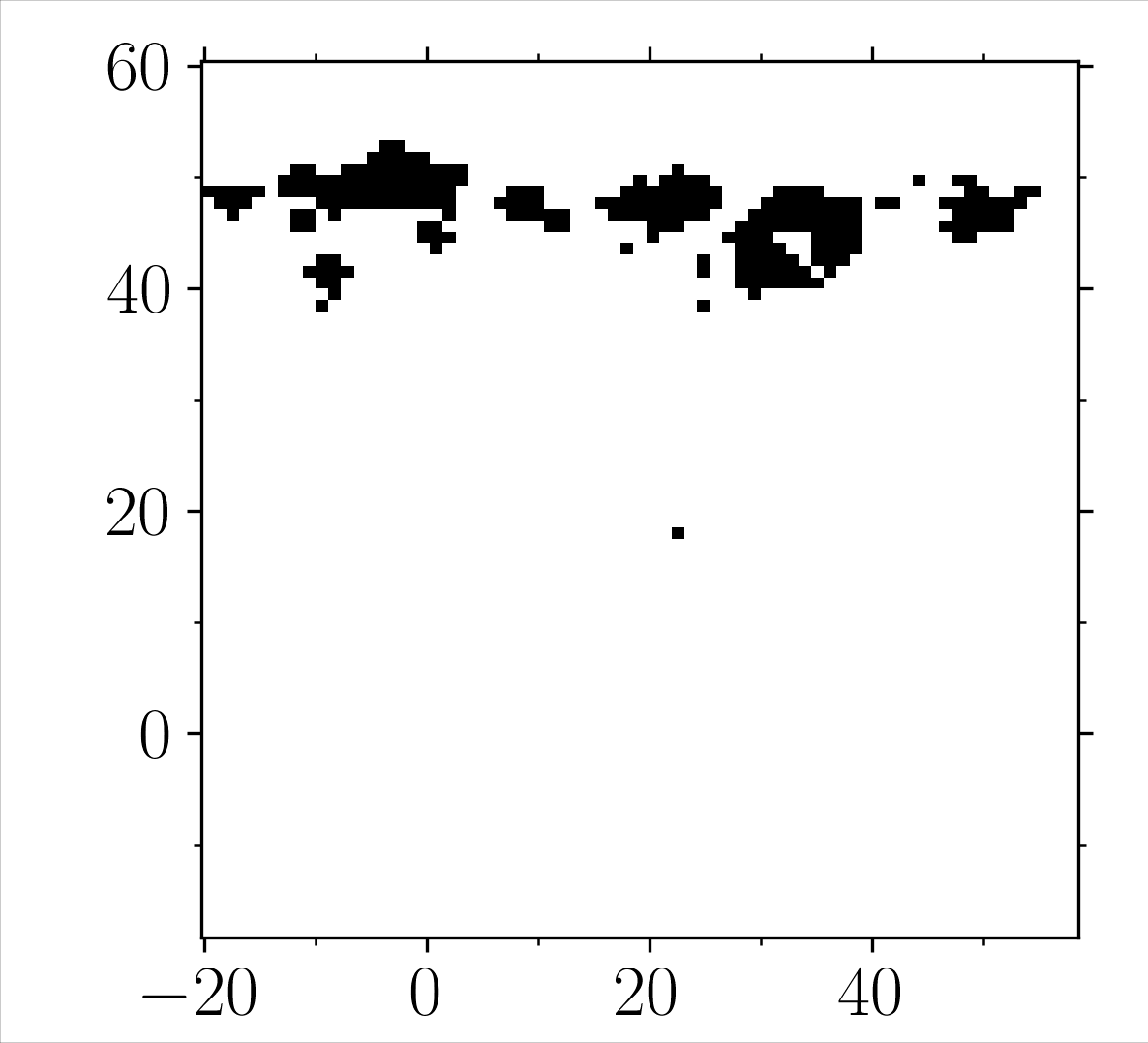}
        \LabelFig{19}{90}{$a)~\gamma_{xy} = 0.1$}
        \Labelxy{46}{-5}{0}{$x$(\r{A})}
        \Labelxy{-6}{44}{90}{$y$(\r{A})}
    \end{overpic}
    \begin{overpic}[width=0.49\columnwidth]{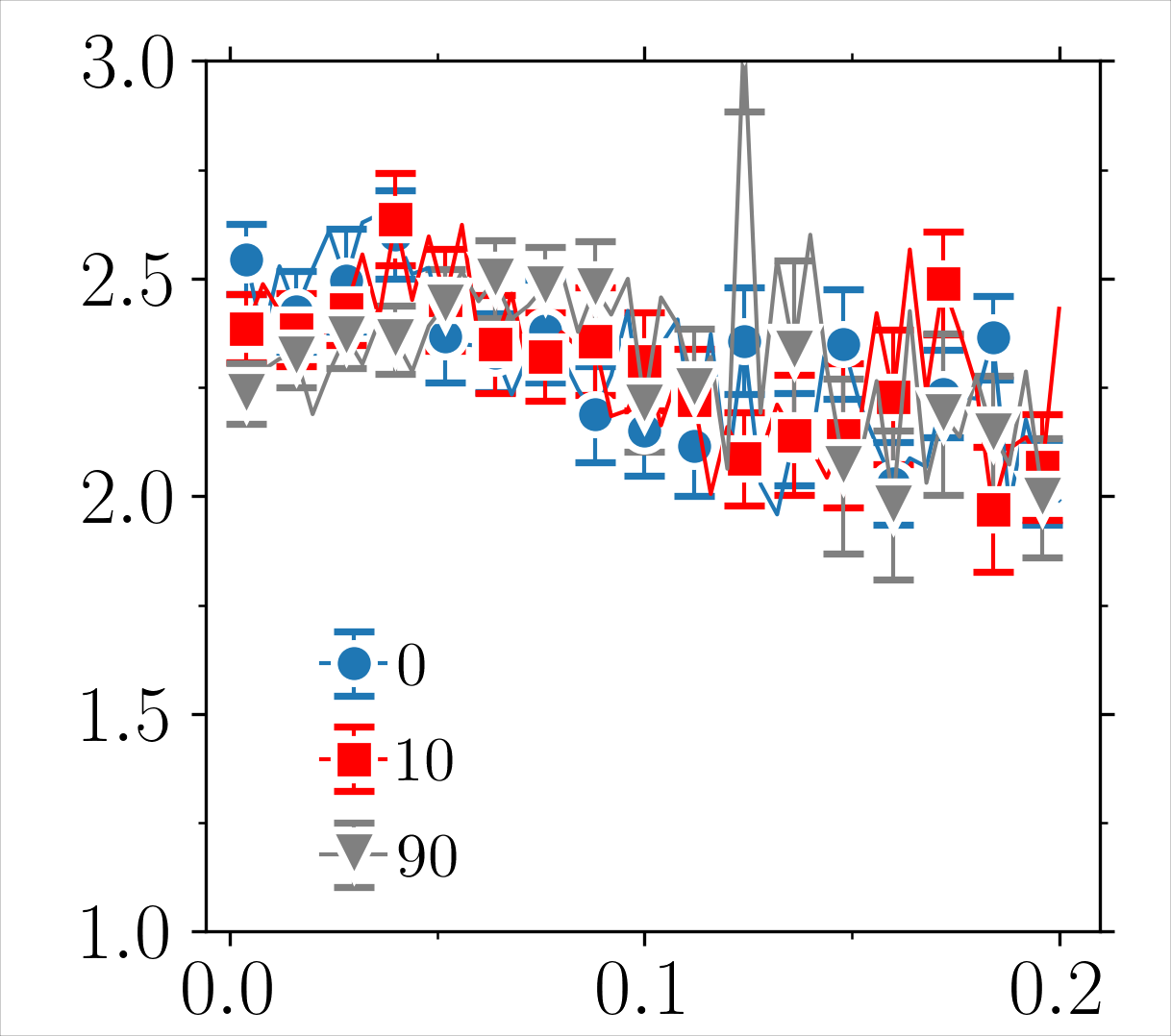}
         \LabelFig{19}{88}{$b)$}
        \Labelxy{50}{-5}{0}{$\gamma_{xy}$}
        \Labelxy{-2}{46}{90}{$d_f$}
        \Labelxy{20}{12}{90}{ $\scriptstyle t_\text{age}(\text{ps})$ }
    \end{overpic}
    % \vspace{-4pt}
    % 
    %
    \begin{overpic}[width=0.49\columnwidth]{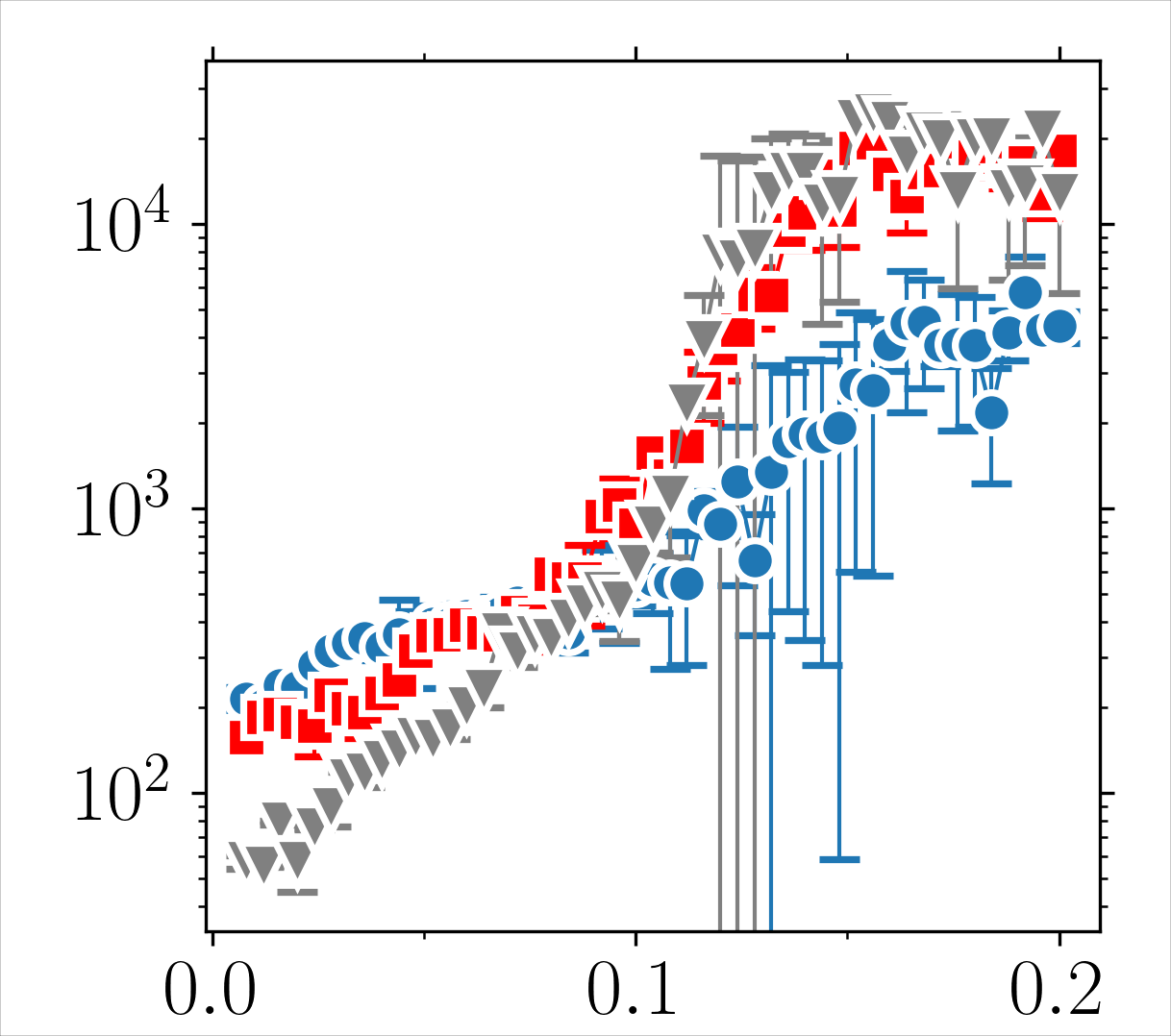}
         \LabelFig{19}{88}{$c)$}
        \Labelxy{50}{-5}{0}{$\gamma_{xy}$}
        \Labelxy{-2}{46}{90}{{$S$\tiny (\r{A}$^3$)}}
    \end{overpic}
    \begin{overpic}[width=0.49\columnwidth]{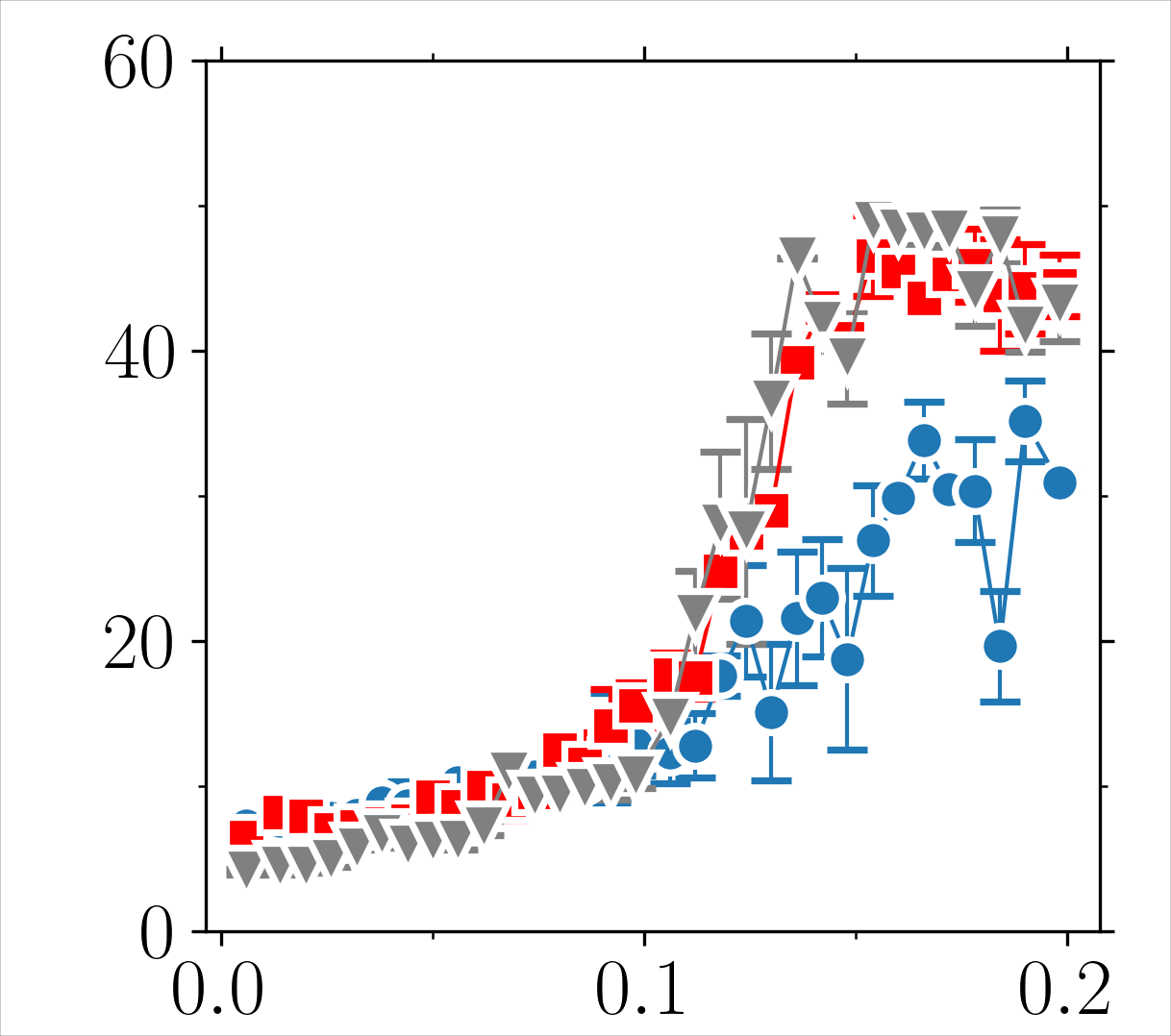}
         \LabelFig{19}{88}{$d)$}
        \Labelxy{50}{-5}{0}{$\gamma_{xy}$}
        \Labelxy{-2}{46}{90}{{$\xi$\tiny (\r{A})}}
    \end{overpic}
    %
%%%%%%%%%%%%%%%%%%%%%%%%%%%%%%%%%%%%%%%%%%%%%%%%%%%%%%%%%%%%%%%%%%%%
    \caption{\textbf{a}) Binary \dmin map at  $\gamma_{xy}=0.1$ corresponding to the annealed \glfour at \tage$=90$ ps. \textbf{b}) Fractal dimension $d_f$ \textbf{c}) Mean cluster size $S$ \textbf{d}) Correlation length $\xi$ plotted against strain $\gamma_{xy}$ at multiple age $t_\text{age}$. Here $x$ and $y$ denote flow and gradient directions, respectively. The binary maps show top $5\%$ sites with largest \dmin in black. }
    \label{fig:crltn}
\end{figure*}

The aging-induced crossover is also manifested in the evolution of $\langle$\dmin\!\!$\rangle$ (averaged over atoms) with \tage as in Fig.~\ref{fig:orderParameters}(c).
Non-affine displacements are low at small strains but exhibit a clear crossover with increasing age at the onset of yielding ($\gamma_{xy} \simeq 0.1$) above which $\langle$\dmin\!\!$\rangle$ grows almost linearly with strain, irrespective of \tage\!\!.
Figure~\ref{fig:orderParameters}(d) displays the scaled standard deviation associated with atoms’ \dmin as a measure of susceptibility.
The overall trend we observe is akin to the behavior illustrated in Fig.~\ref{fig:orderParameters}(c). 
In this context, fluctuations tend to reveal an age-dependent shear-induced transition that displays characteristics akin to critical phenomena and a diverging correlation length (\cf Fig.~\ref{fig:orderParameters}(d)). The character/order of this shear-induced transition and the exponents associated to this correlation length divergence may be unveiled through finite-size scaling studies in the strong aging regime, a study that goes beyond the purpose of the current work.
% Figure~\ref{fig:pdfCondD2min}(c) and (d) illustrate \dmin probability distribution function corresponding to atoms with icosahedral symmetry in the quenched and annealed sample, respectively.
% The latter exhibits a clear bimodal behavior in Fig.~\ref{fig:pdfCondD2min}(d) with the first (higher) and second (lower) peaks denoting the population of atoms outside and within shear zones (cf. Fig.~\ref{fig:scatterD2minRho}).
% The conditional distribution (red squares) indicates a relatively higher contribution of the ordered icosahedral phase to the higher peak in very close agreement with the previous observation regarding the rare occurrence of ordering within rearranging zones.
% Such features are also present in Fig.~\ref{fig:pdfCondD2min}(c) corresponding to the quenched glass but with less pronounced bimodality.
We also probed variations in the (mean) number density of ordered clusters $\langle$\rhoico\!$\rangle$ with strain as in Fig.~\ref{fig:orderParameters}(b).
Our data show that the degree of ordering tends to increase with  aging,  and the applied shear appears to  further \emph{amorphize/rejuvenate} the deforming glass (\eg by reduction in the icosahedral cluster density by $>10\%$). 
Nevertheless, this observable, lacks a clear signature of the yielding transition and associated variations with age.

The \dmin maps depicted in Fig.~\ref{fig:scatterD2minRho} give a visual impression that the  elastic-plastic transition in BMGs (at $\gamma_{xy} \simeq 0.1$) might indeed coincide with a \emph{percolation} transition of rearranging atoms upon shear loading (see also \cite{karimi2023yielding} and references therein).
To validate this picture, we adopt ideas from classical percolation theory \cite{stauffer2018introduction}, including investigations of cluster sizes and their dynamical evolution. 
To perform the cluster analysis, atom-wise \dmin was interpolated on a regular (cubic) grid and the top $5\%$ of grid points with highest \dmin were labeled as rearranging sites and colored in black in Fig.~\ref{fig:crltn}(a).
As a basic statistical property, $n_s$ denotes the probability distribution function associated with the number of clusters containing $s$ rearranging sites.
The radius of gyration associated with a cluster of size $s$ is defined as $r^2_s=\sum_{i=1}^{s}|\vec{r}_i-\vec{r}_0|^2/s$ with the center of mass $\vec{r}_0=\sum_{i=1}^{s}\vec{r_i}/s$.
We obtain $s\propto r_s^{d_f}$ with fractal dimension $d_f$.  
The mean cluster size is defined as $S={\sum_s n_ss^2}/{\sum_s n_ss}$.
We also define the (squared) correlation length $\xi^2={2\sum_s r^2_ss^2n_s}/{\sum_ss^2n_s}$, based on a weighted average that is associated with the radius of gyration $r^2_s=\sum_{i=1}^{s}|\vec{r}_i-\vec{r}_0|^2/s$ of a cluster of size $s$.
%Here the center of mass is $\vec{r}_0=\sum_{i=1}^{s}\vec{r_i}/s$.

The evolution of fractal dimension $d_f$ with strain is illustrated in Fig.~\ref{fig:crltn}(b),  corresponding to three different sample ages.
The overall reduction towards $d_f=2$ may imply that the soft spots tend to form fairly compact clusters at low strains but favor a more planar topology on approach to yielding.   
Displayed in Fig.~\ref{fig:crltn}(c) and (d), both mean cluster size $S$ and correlation length $\xi$ indicate a fairly smooth evolution with strain at \tage$=0$ but develop quite sharp features as the age is increased towards \tage$=90$ps.
The correlation length tends to saturate at $\xi \simeq 45$ \r{A} due to the physical size limit, in a visual agreement with cluster maps illustrated in Fig.~\ref{fig:crltn}(a) and (b). 
The overall reduction in size with increasing \tage at initial stages of deformation indicates that annealing and associated structural relaxation leads to annihilation of STZs, in contrast to the observed enhancement in the density of SROs (cf. Fig.~\ref{fig:orderParameters}(b)). 
% Our work has mainly focused on composition-dependent ductility and its correlations with softness percolation.
% Such correlations are potentially analogous to the preparation effects (i.e. quenching versus annealing) on the nature of yielding transition.
% Fast quench rates, on the other hand, tend to further amorphize (or \emph{rejuvenate}) the glassy structure culminating in denser populations of STZs and, thus, enhanced ductility and (fracture) toughness.  
% The investigation of the annealing protocol and its effects on softness-yielding correlations, as evidenced in Fig.~\ref{fig:hmin_gamma}, is an interesting topic that could be reserved as future work.
\\

\emph{Conclusions~--~}
The present study of sheared  BMGs has brought new insights into the underlying correlations between aging, structural ordering, and strain localization. 
We have presented direct evidence that the annealing process plays a pivotal role in controlling the sharpness of the shear-induced elastic-plastic transition, leading to a crossover from diffuse deformation features in as-quenched samples to localized shear-band-like patterns in well-annealed deforming glasses. 
%This phenomenon is reminiscent of the ductile-to-brittle transition observed in various solid materials.
Our findings suggest that the observed crossover is rooted in the interplay between aging-induced icosahedral ordering and collective formation of STZs.
This observation has been quantified via probing several order parameters coupled with measurable dynamical and structural metrics. 
This includes fluctuations in the atoms' propensity to plastic rearrangements as well as spatial variations in the local density of icosahedral clusters.
By analyzing connected networks of soft rearranging regions, we extracted relevant length scales that evolve across the yielding transition and exhibit significant variations with the sample's age.
% Furthermore, we observed a pronounced stress overshoot in the load curves, which strongly depends on the glass age. This stress overshoot is associated with the initiation of shear bands and serves as a robust indicator of strain localization. The evolution of the mean squared nonaffine displacements (\dmin) and its correlation with icosahedral ordering density (\rhoico) demonstrated that the degree of structural ordering is inversely related to strain localization within shear bands.
Our findings contribute to a better understanding of the complex interplay between structural order, aging, and plasticity in metallic glasses. This knowledge has implications for the design and optimization of metallic glasses for various engineering applications, whereas the control over strain localization and ductility via preparation protocols may be crucial.

%\add[KK]{The present study brings new insights about i) the microstructure-property paradigm in designing ductile multicomponents bulk metallic glasses ii) the notion of elastic heterogeneity as a robust micro-mechanical indicator of plasticity. Along the lines of i), the percolation of local elasticity upon failure and its precursory nature connects directly with the previous percolation studies. We, however, have revisited this concept within the framework of composition-dependence of yielding transition and probed its meaningful variations/correlations with softness percolation. As for ii), our work complements ongoing efforts within the glass community that aim to predict plasticity based on the notion of non-affinity. Such predictions, mainly centered on structural/topological signatures of failure precursors, can be substantially improved by bringing micro-mechanical aspects (e.g. elastic heterogeneity) into the picture.}

\emph{Acknowledgments~--~}
This research was funded by the European Union Horizon 2020 research and innovation program under grant agreement no. 857470 and from the European Regional Development Fund via Foundation for Polish Science International Research Agenda PLUS program grant no. MAB PLUS/2018/8.
%We wish to thank A. Esfandiarpour and R. Alvarez for providing the data sets.
%\end{acknowledgments}

\bibliography{references}% Produces the bibliography via BibTeX.

\end{document}

\title{Tuning brittleness in multi-component metallic glasses through chemical disorder aging}% Force line breaks with \\

\author{Kamran Karimi$^1$}
%\email{Corresponding author: kamran.karimi@ncbj.gov.pl}
%\author{Amin Esfandiarpour$^1$}%
%\author{Ren\'e Alvarez-Donado$^1$}
% \author{Mikko J. Alava$^{1,2}$}
\author{Stefanos Papanikolaou$^1$}
\affiliation{%
 $^1$ NOMATEN Centre of Excellence, National Center for Nuclear Research, ul. A. Sołtana 7, 05-400 Swierk/Otwock, Poland\\
% $^{2}$ Aalto University, Department of Applied Physics, PO Box 11000, 00076 Aalto, Espoo, Finland
}%

\maketitle

\section*{Supplementary Materials}
In this Supplementary Materials, we will provide additional details relevant to the preparation protocols of the \glfour\!\!, \glfive\!\!, \gltwo\!\!, and \glone metallic glasses presented in this study.

\subsection*{Simulations and Protocols}\label{sec:SimulationsAndProtocols}
Molecular dynamics simulations were carried out in LAMMPS \cite{plimpton1995fast} by implementing atomistic systems of size $N=50,688$ within periodic cubic boxes with $L\simeq 80$ \r{A} in a simple-shear loading geometry.
The above compositions are characterized by fairly low misfit coefficients $\delta_a$ and, therefore, low glass forming ability owing to effects of atomic size distributions. 
Nonetheless, there is a way to generate a model glass in the opposite limit of $\delta_a \rightarrow 0$ by maintaining a larger number of elements, as in medium/high entropy metallic glasses and/or using higher cooling rates (see \cite{ding2013high,alvarez2023viewing} and references therein).

The interatomic forces were derived from the modified embedded atom method (MEAM) potential with the input parameters associated with the principle elements obtained from \cite{choi2018understanding}.
The glassy samples were prepared based on the melt-and-quench protocol where a crystalline structure was initially heated up to a high temperature ($3000$~K) above the melting temperature $T>T_m$, before it was cooled down to the room temperature ($300$~K) below $T_g$ at the quench rate of $\dot{T}= 0.1$~K/ps.
We allowed for further relaxation upon quenching for order $10$ ps to ensure as-quenched glasses are fully equilibrated prior to shearing.
The ambient pressure $P_0$ was set to zero at all times.
We also set the discretization time to $\Delta t=0.001$ ps.
The NPT ensembles were implemented via a Nose-Hoover thermostat and barostat with relaxation time scales $\tau_d^\text{therm}=0.2$~ps and $\tau_d^\text{bar}=2.5$~ps. 

\begin{figure*}[t]
    \centering
    \begin{overpic}[width=0.49\columnwidth]{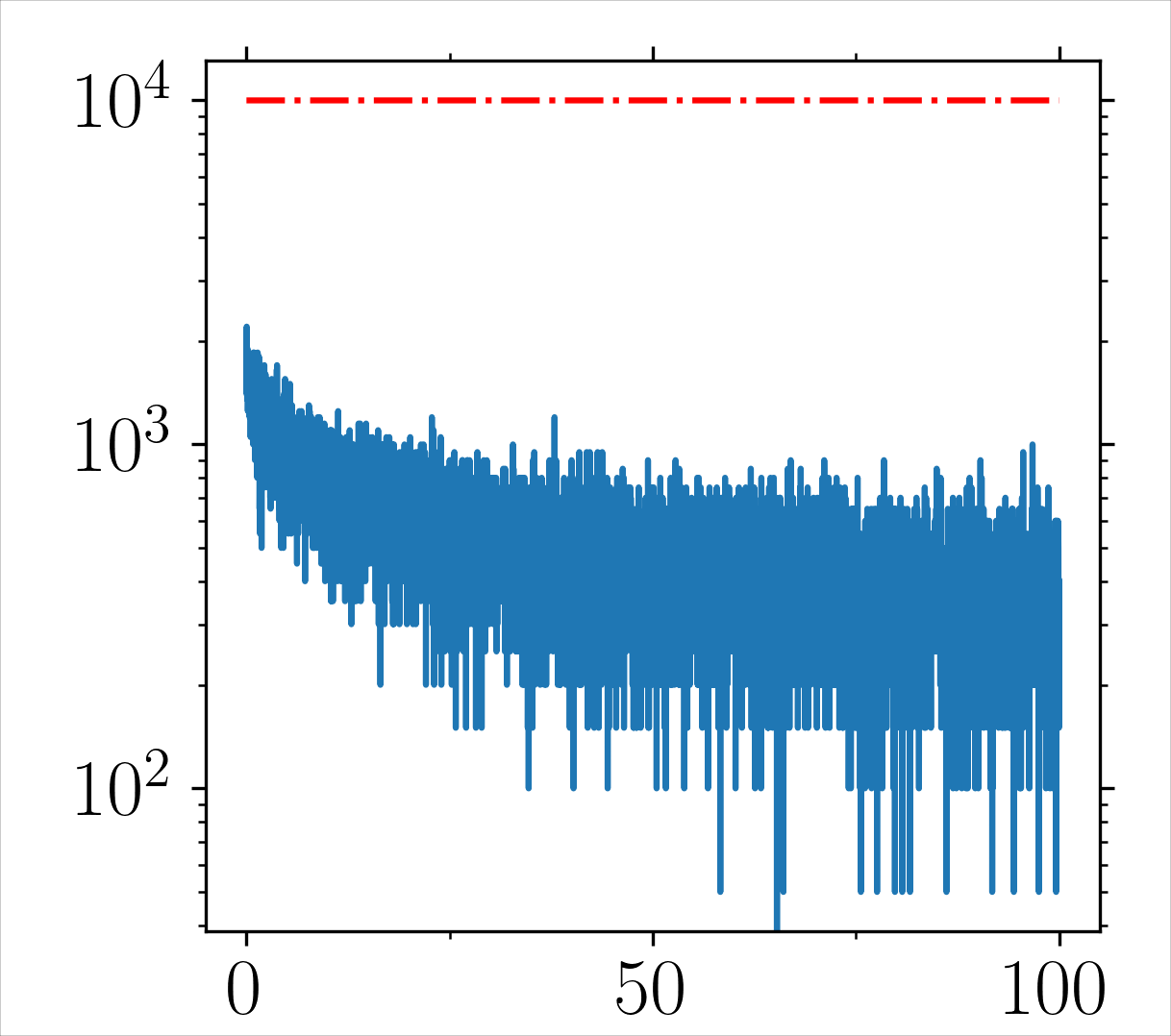}
        \LabelFig{19}{88}{$a)$}
        \Labelxy{50}{-5}{0}{\tage (ps)}
        \Labelxy{-2}{30}{90}{Swap Rate \tiny(ps$^{-1}$)}
      \end{overpic}
    \begin{overpic}[width=0.49\columnwidth]{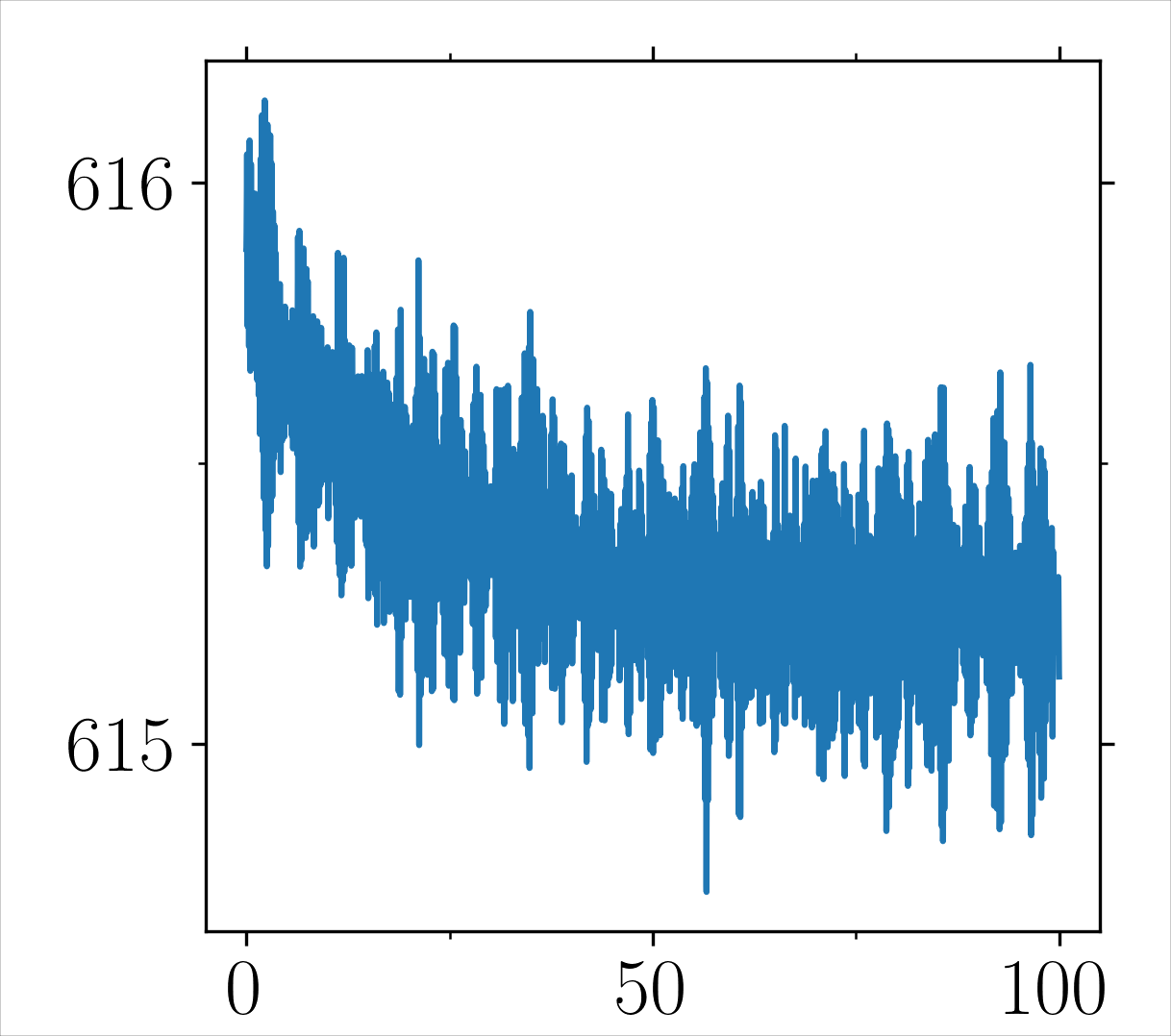}
        \LabelFig{19}{88}{$b)$}
        \Labelxy{50}{-5}{0}{\tage (ps)}
        \Labelxy{-2}{30}{90}{Volume \tiny ($\times 10^3$\r{A}$^3$)}
      \end{overpic}
    \begin{overpic}[width=0.49\columnwidth]{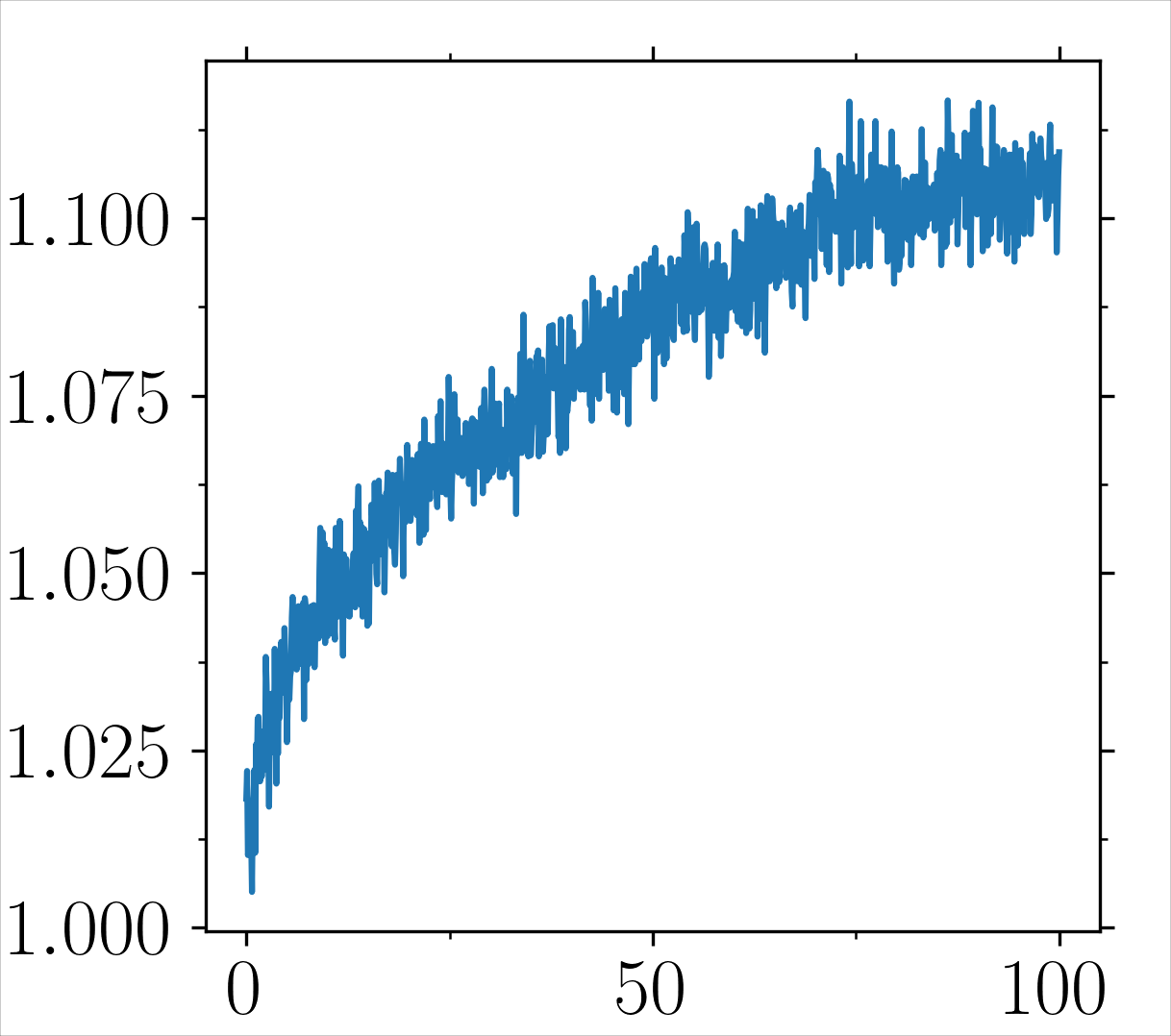}
        \LabelFig{19}{88}{$c)$}
        \Labelxy{50}{-5}{0}{\tage (ps)}
        \Labelxy{-8}{30}{90}{$\langle$\rhoico\!$\rangle$ \tiny(\r{A}$^{-3}$)}
      \end{overpic}
    \caption{\textbf{a}) Accepted swapping rate \textbf{b}) volume \textbf{c}) mean number density of atoms with icosahedral symmetry $\langle$\rhoico\!$\rangle$ plotted against aging duration $t_\text{age}$ in \glfive\!\!. The red flat line indicates the attempted swapping rate.}
    \label{fig:thermo}
\end{figure*}

\begin{figure*}[t]
    \centering
    \begin{overpic}[width=0.49\columnwidth]{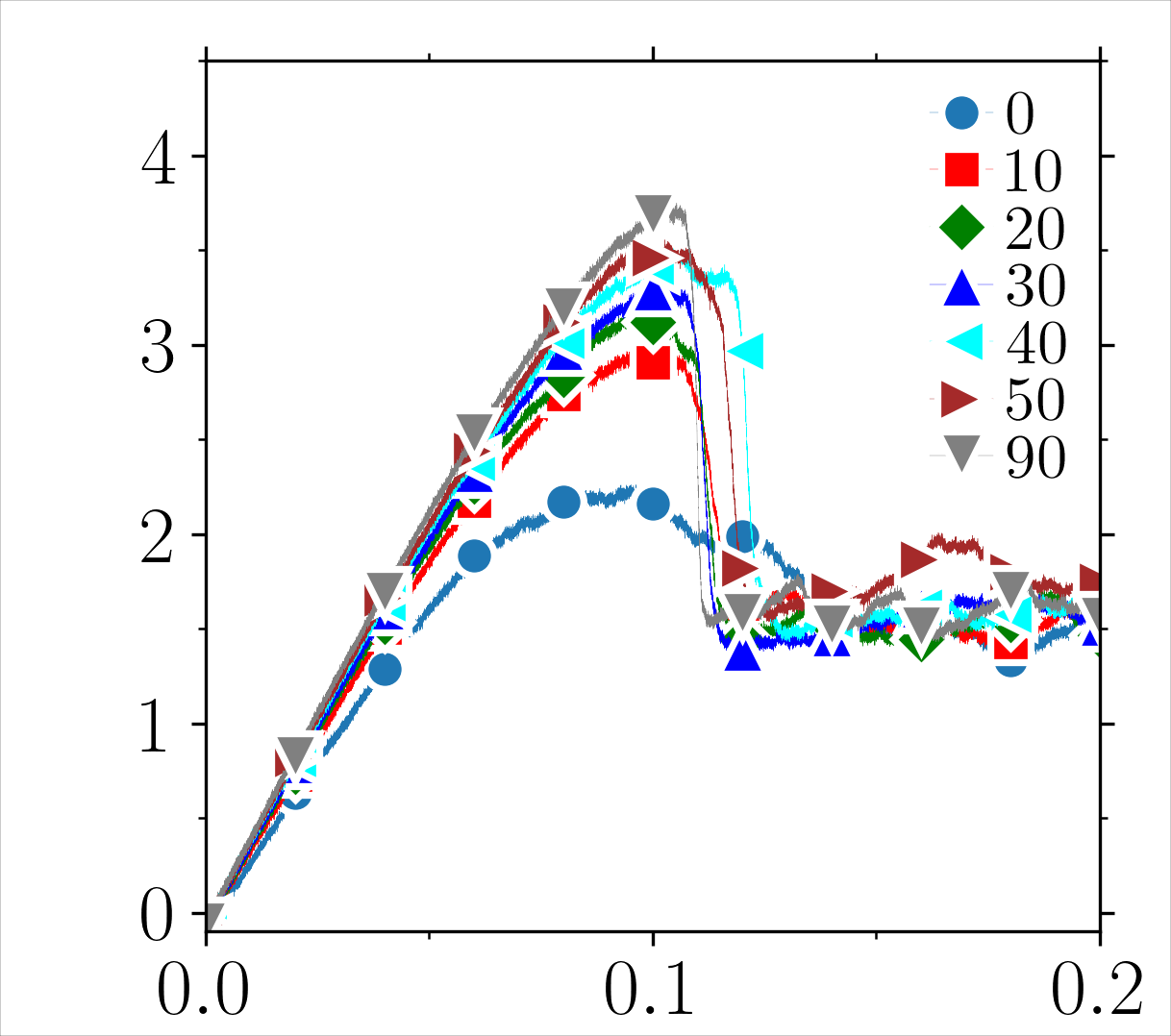}
        \LabelFig{19}{88}{$a)$ \scriptsize \glfive}
        \Labelxy{50}{-5}{0}{$\gamma_{xy}$}
        \Labelxy{3}{35}{90}{{$\sigma_{xy}$ \tiny(Gpa)}}
        \Labelxy{72}{49}{90}{ $\scriptstyle t_\text{age}(\text{ps})$ }
      \end{overpic}
    \begin{overpic}[width=0.49\columnwidth]{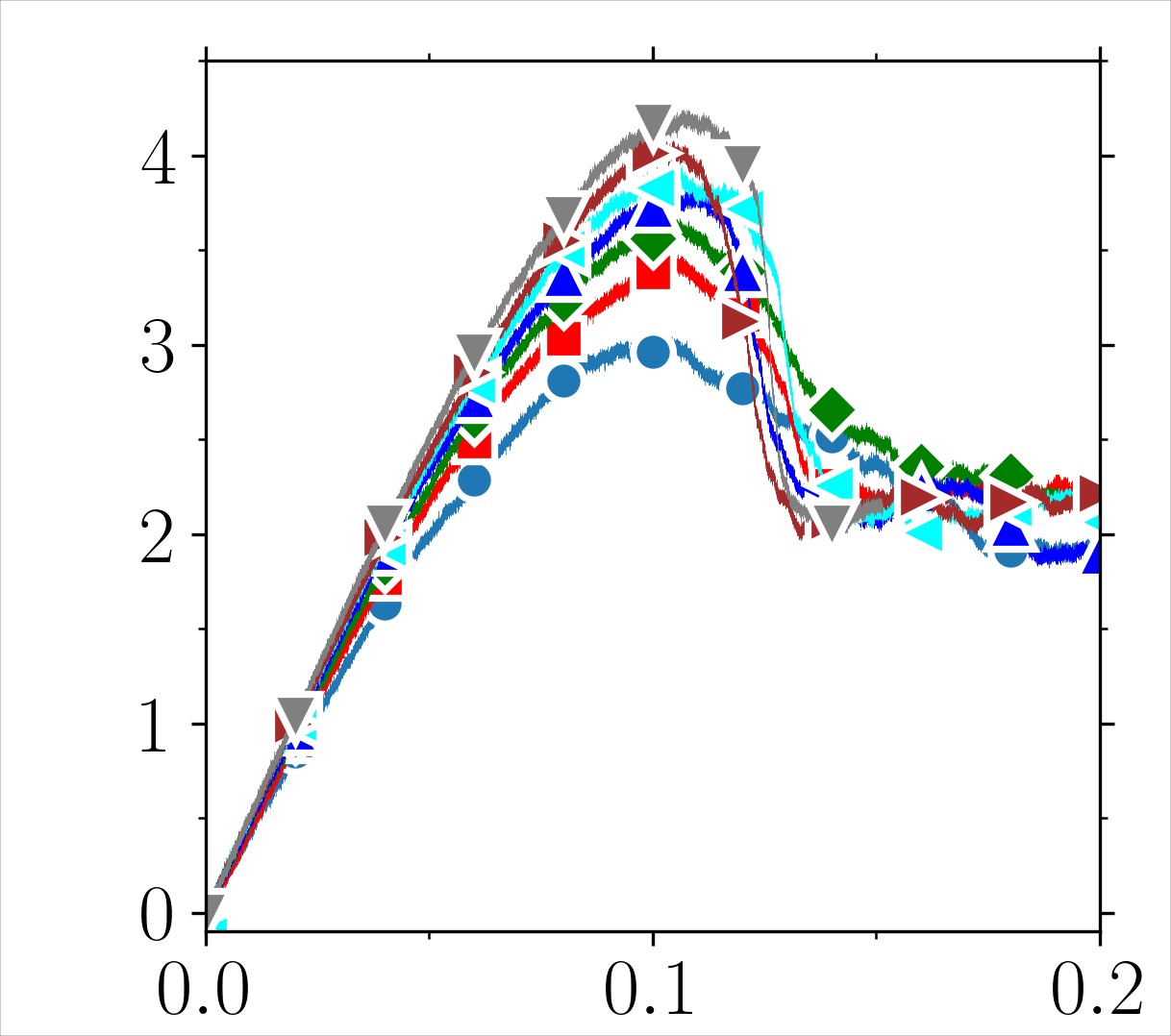}
        \LabelFig{19}{88}{$b)$ \scriptsize \gltwo}
        \Labelxy{50}{-5}{0}{$\gamma_{xy}$}
        \Labelxy{3}{35}{90}{{$\sigma_{xy}$ \tiny(Gpa)}}
      \end{overpic}
     %
%    \begin{overpic}[width=0.49\columnwidth]{Figs2nd/stress_strain_age_conicrfe.png}
        %
%        \LabelFig{19}{88}{$c)$ \scriptsize \glone}
        %
%        \Labelxy{50}{-5}{0}{$\gamma_{xy}$}
%        \Labelxy{3}{35}{90}{{$\sigma_{xy}$ \tiny(Gpa)}}
%      \end{overpic}
     %
    \caption{Macroscopic stress $\sigma_{xy}$ plotted against applied (shear) strain $\gamma_{xy}$ at several aging duration $t_\text{age}$ corresponding to \textbf{a}) \glfive \textbf{b}) \gltwo.}
    \label{fig:loadCurve_compositions}
\end{figure*}

\begin{figure*}[t]
    \centering
    \begin{overpic}[width=0.49\columnwidth]{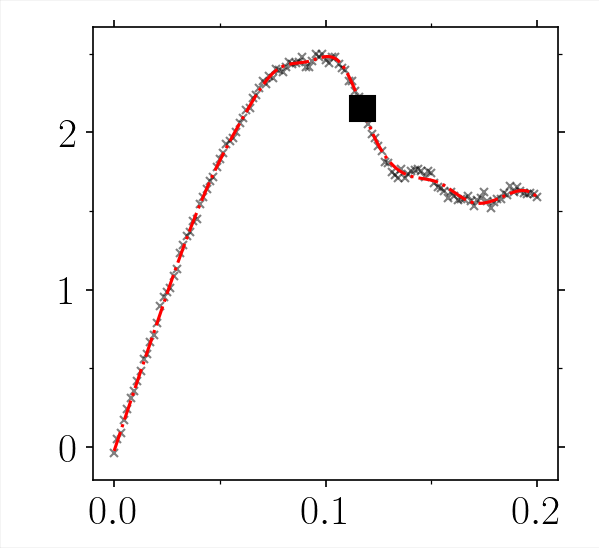}
%        \LabelFig{17}{81}{$(a)$}
        \Labelxy{50}{-3}{0}{$\gamma_{xy}$}
        \Labelxy{-2}{35}{90}{$\sigma_{xy}$ \scriptsize (Gpa)}
       \begin{tikzpicture}
            \coordinate (a) at (0,0); % reference
            \node[white] at (a) {\tiny.};               %   
                % draw slope
%            \drawSlope{3.0}{3.4}{0.4}{20}{black}{\hspace{-8pt}\large h_\text{min}}
            \drawSlope{2.35}{2.7}{0.4}{25}{black}{\hspace{-5pt}$\scriptstyle h_\text{min}$}
            % draw arrow
			\coordinate (center2) at (2.45,2.9); % center of figure
            \coordinate (b) at ($ (center2) - .2*(1,-2.3) $); % length=0.4
            \coordinate (c) at ($ (center2) + .2*(1,-2.3) $);
            \draw[black][line width=0.4mm] (c) -- (b); % horizontal line
            \coordinate (d) at ($ (center2) - .5*(0,1) $); %arrow length
		    \end{tikzpicture}
		    
	\end{overpic}
    \caption{Illustration of the softening modulus \hmin. The (black) crosses indicate the stress data points $\sigma_{xy}$ as a function of the applied (shear) strain $\gamma_{xy}$. The dashdotted (red) curve denotes a nonlinear fit based on (cubic) smoothing splines.  The square symbol indicates the strain corresponding to the minimal slope past the stress peak as illustrated by the straight line.}
    \label{fig:hminIllustration}
\end{figure*}

The annealing process was implemented through Monte Carlo swaps between atoms of different types at a specified (annealing) temperature $T_a=300$ K, the (attempted) swapping rate of $10^4$ pairs per ps, and over the aging duration of $t_\text{age}=100$ ps.
Figure~\ref{fig:thermo} plots the accepted swapping rate, volume, and mean number density of atoms with icosahedral symmetry $\langle$\rhoico\!$\rangle$ as a function of annealing time $t_\text{age}$ in \glfive\!\!.
%Prior to shearing, the bulk stress tensor $\sigma_{\alpha\beta}\simeq-P_0~ \delta_{\alpha\beta}$ has a nearly hydrostatic form.
A strain-controlled condition was applied by deforming the periodic box on the $xy$ plane implementing Lees-Edwards boundary conditions \cite{lees1972computer} at a fixed volume and temperature (canonical NVT ensembles) and a constant shearing rate $\dot\gamma_{xy}=10^{-4}$ ps$^{-1}$ up to a $20\%$ strain.
Figure~\ref{fig:loadCurve_compositions} displays macroscopic stress $\sigma_{xy}$ against applied (shear) strain $\gamma_{xy}$ at several aging duration $t_\text{age}$ corresponding to \glfive\!\!, \gltwo\!\!, and \glone\!\!.
%During the loading phase, the simulation box remained coupled to the barostat along $x$ with $\sigma_{xx}$ retained at the pre-compression level $-p_0$.
%The numerical range of the applied pressure $p_0$ we have tested corresponds to the average overlap $\frac{\delta}{R_s}\simeq0.01-0.1$.
%We also checked that $\dot\gamma$ was slow enough leading to a stress evolution that was almost insensitive to the loading rate (the so-called ``rate independent" regime) \note[KK]{did we?}.
%Within this rate-independent regime, the slowly driven solid reveals contrasting failure patterns depending on $\mu$.
%The results of the compression test may be used to determine the bulk shearing strength together with the structure of shear deformations approaching failure.
We further measure the softening modulus \hmin defined as the maximum \emph{rate} of the \emph{macroscopic} stress drop at every age \tage as in Fig.~\ref{fig:hminIllustration}.

%%%%%%%%%%%%%%%%%%%%%%%%%%%%%%%%%%%%%%%%%%%%%%%%%%%%%%%%%%%%%%%%%%%%
%%%%%%%%%%% figure
%%%%%%%%%%%%%%%%%%%%%%%%%%%%%%%%%%%%%%%%%%%%%%%%%%%%%%%%%%%%%%%%%%%%        

%%%%%%%%%%%%%%%%%%%%%%%%%%%%%%%%%%%%%%%%%%%%%%%%%%%%%%%%%%%%%%%%%%%%
%%%%%%%%%%% figure
%%%%%%%%%%%%%%%%%%%%%%%%%%%%%%%%%%%%%%%%%%%%%%%%%%%%%%%%%%%%%%%%%%%%        

%%%%%%%%%%%%%%%%%%%%%%%%%%%%%%%%%%%%%%%%%%%%%%%%%%%%%%%%%%%%%%%%%%%%
%%%%%%%%%%% figure
%%%%%%%%%%%%%%%%%%%%%%%%%%%%%%%%%%%%%%%%%%%%%%%%%%%%%%%%%%%%%%%%%%%%

%%%%%%%%%%%%%%%%%%%%%%%%%%%%%%%%%%%%%%%%%%%%%%%%%%%%%%%%%%%%%%%%%%%%
%%%%%%%%%%% figure
%%%%%%%%%%%%%%%%%%%%%%%%%%%%%%%%%%%%%%%%%%%%%%%%%%%%%%%%%%%%%%%%%%%%
\begin{figure*}[b]
%    \raggedright
     \begin{overpic}[width=0.49\columnwidth]{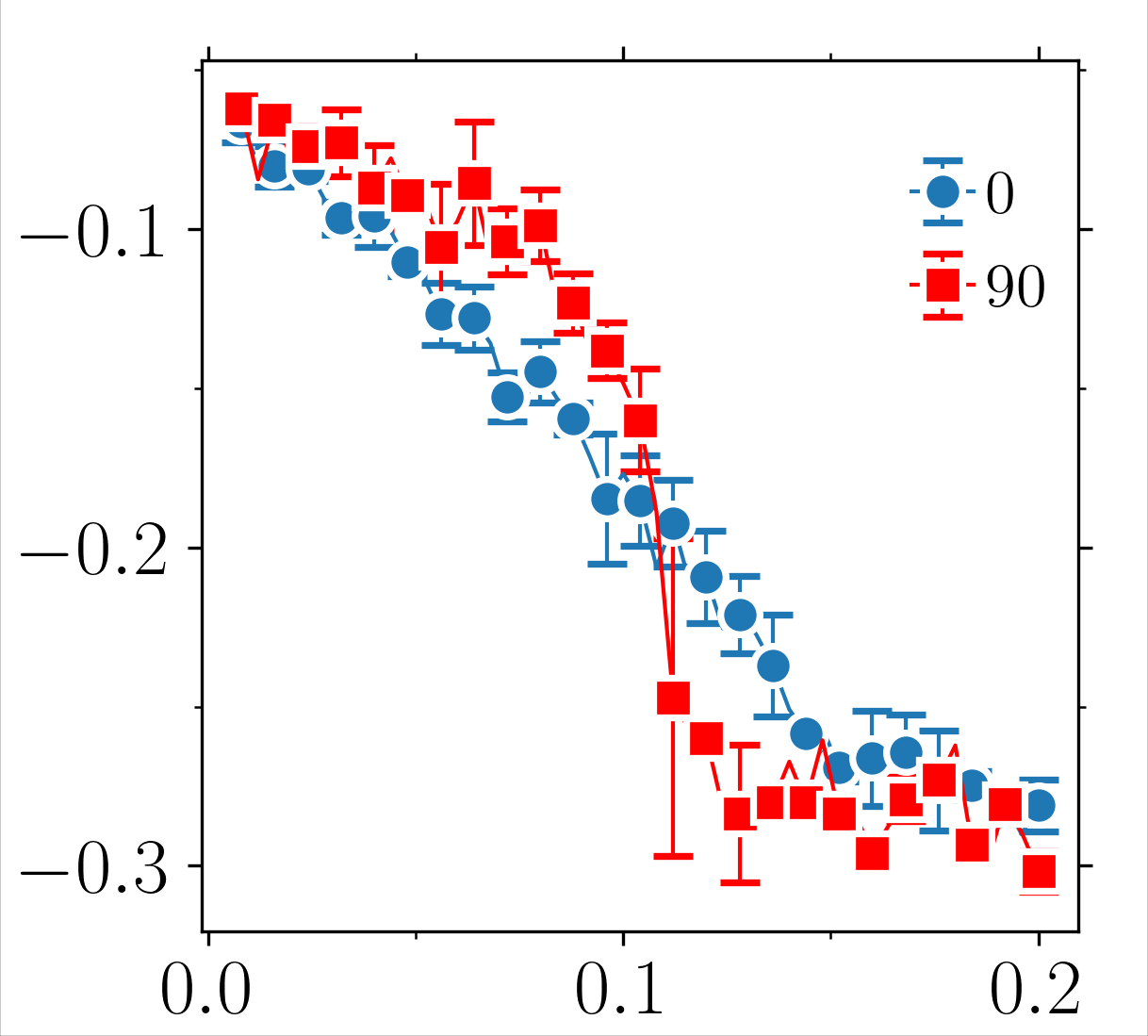}
        \Labelxy{50}{-3}{0}{$\gamma_{xy}$}
        \Labelxy{-4}{35}{90}{$c_{XY}$}
        \Labelxy{72}{56}{90}{ $\scriptstyle t_\text{age}(\text{ps})$ }
        \LabelFig{19}{76}{$a)$}
    \end{overpic}
    \begin{overpic}[width=0.49\columnwidth]{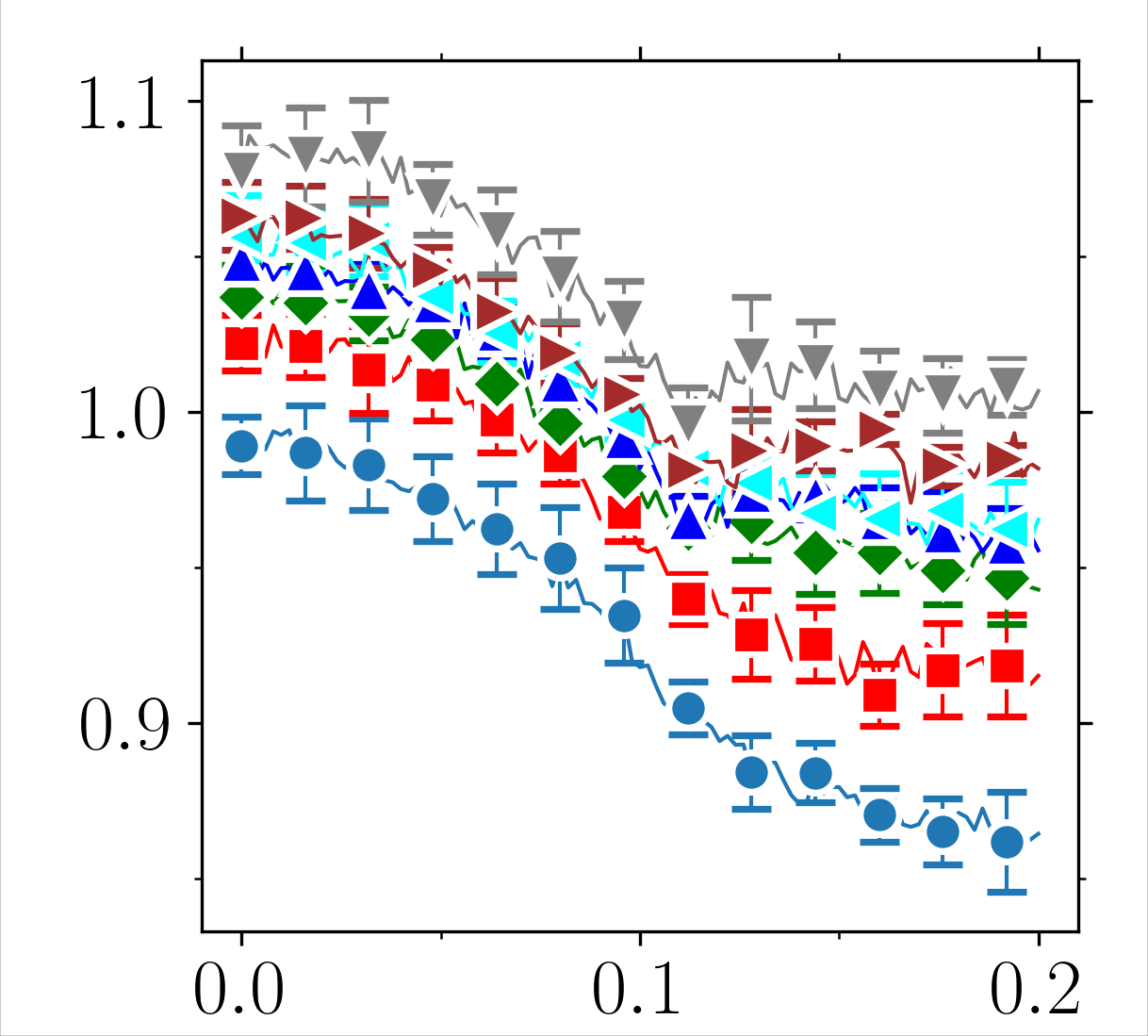}
        \LabelFig{19}{76}{$b)$}
        \Labelxy{50}{-3}{0}{$\gamma_{xy}$}
%        \Labelxy{-2}{32}{90}{$\langle$ \dmin $\rangle$ \tiny(\r{A}$^2$)}
        \Labelxy{-5}{32}{90}{$\langle$\rhoico\!$\rangle$ \tiny(\r{A}$^{-3}$)}
        % \put(100,30){\includegraphics[width=0.07\columnwidth]{Figs2nd/rhoIco_strain_age_legend.png}}  
        % \Labelxy{100}{83}{0}{ $\scriptstyle t_\text{age}(\text{ps})$ }
    \end{overpic}
    \begin{overpic}[width=0.49\columnwidth]{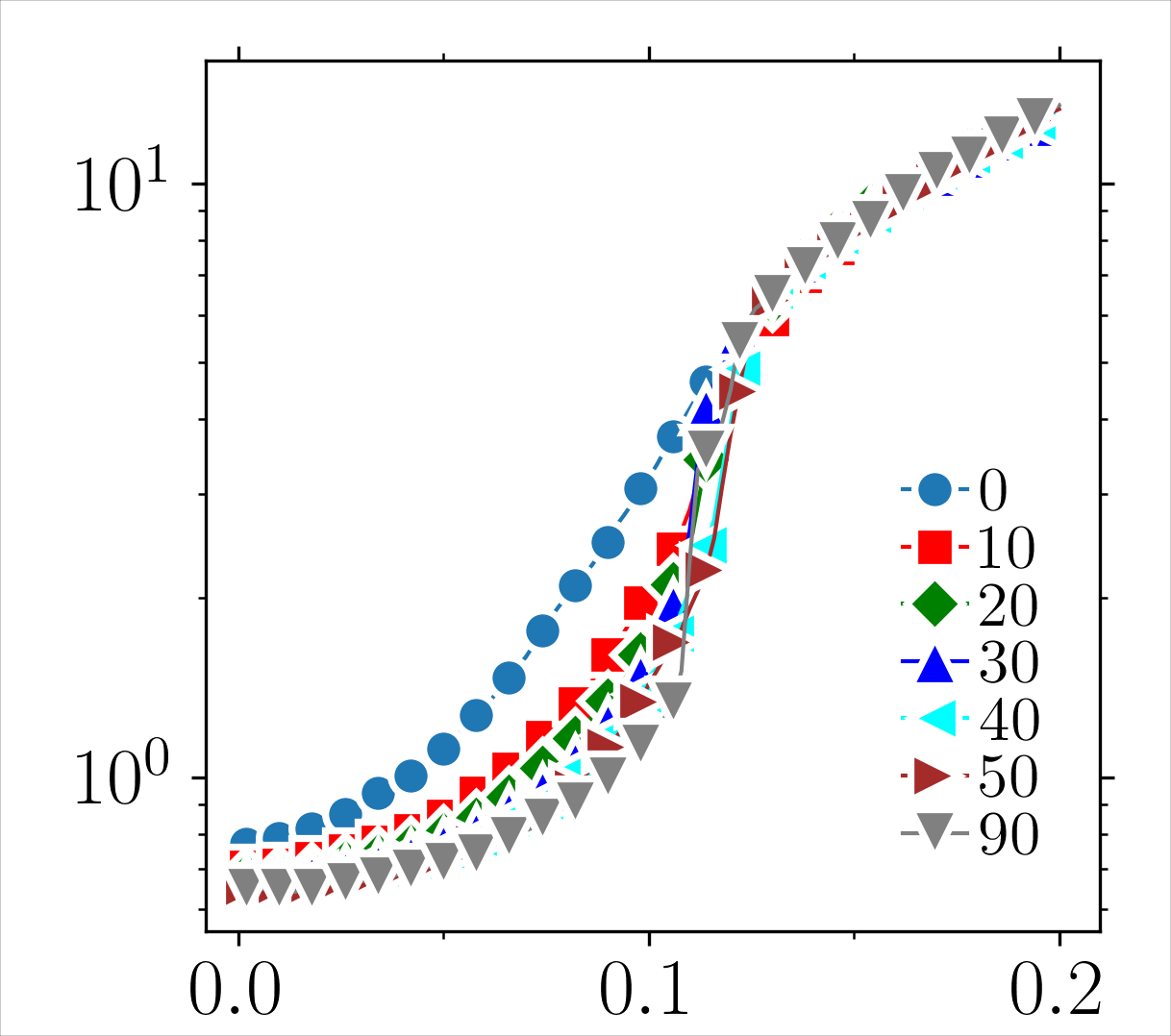}
        \Labelxy{50}{-4}{0}{$\gamma_{xy}$}
        \Labelxy{-2}{32}{90}{$\langle$\dmin\!\!$\rangle$ \tiny(\r{A}$^2$)}
        \Labelxy{69}{16}{90}{ $\scriptstyle t_\text{age}(\text{ps})$ }
         \LabelFig{19}{76}{$c)$}
    \end{overpic}
    \begin{overpic}[width=0.49\columnwidth]{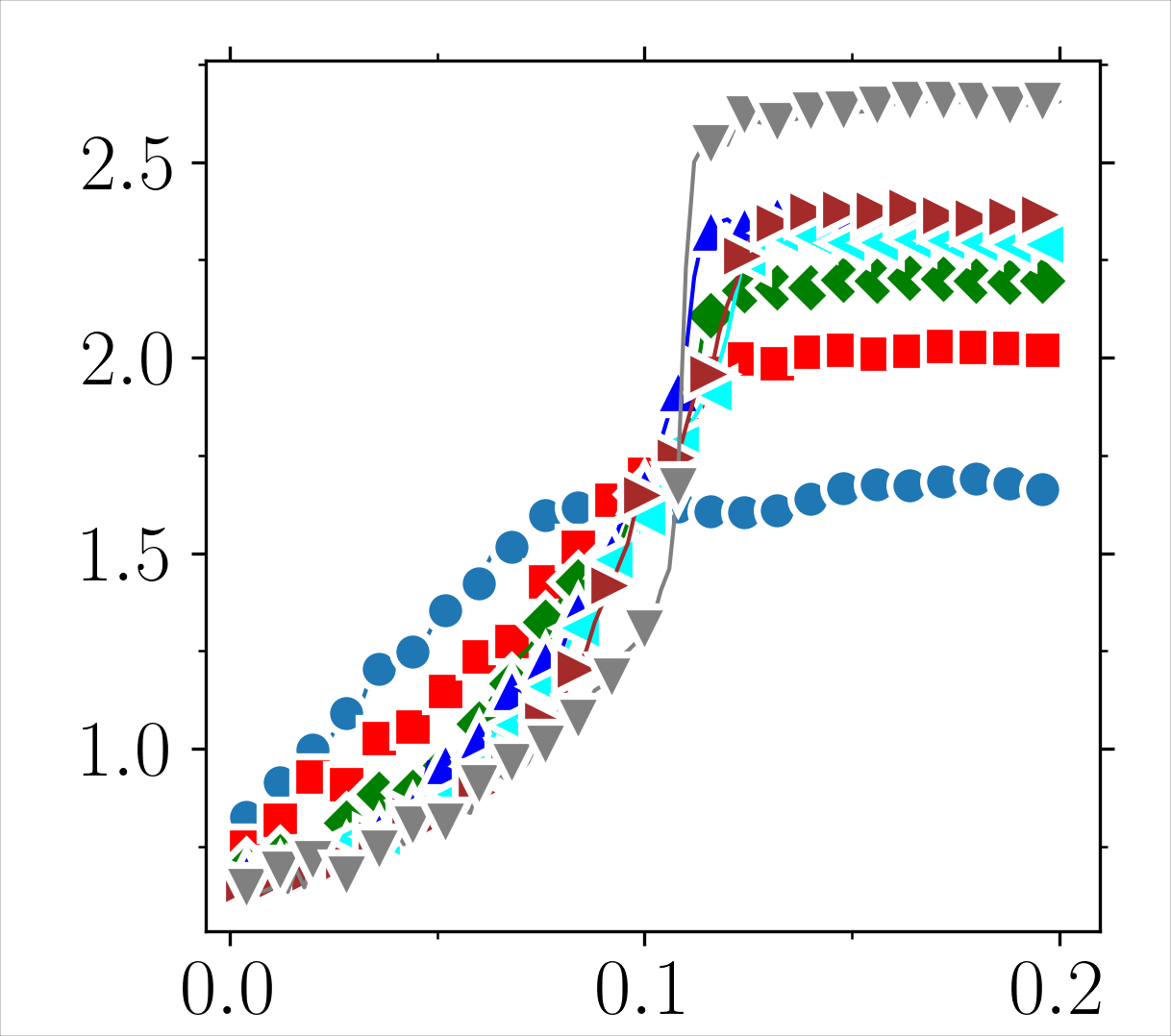}
        \Labelxy{50}{-4}{0}{$\gamma_{xy}$}
        \Labelxy{-5}{20}{90}{std(\dmin\!)/$\langle$\dmin\!\!$\rangle$}
         \LabelFig{19}{76}{$d )$}
    \end{overpic}
    \caption{Evolution of \textbf{a}) correlation coefficient $c_{XY}$ \textbf{b}) mean number density of atoms with icosahedral symmetries $\langle\rho^\text{ico}\rangle$ \textbf{c}) mean squared nonaffine displacements $\langle$\dmin\!\!$\rangle$ \textbf{d}) scaled standard deviation std(\dmin\!)/$\langle$\dmin\!\!$\rangle$ associated with atoms' \dmin plotted against applied strain $\gamma_{xy}$ in \glfive corresponding to different aging duration $t_\text{age}$. The error bars denote standard errors.}
    \label{fig:orderParametersGlassFive}
\end{figure*}

%%%%%%%%%%%%%%%%%%%%%%%%%%%%%%%%%%%%%%%%%%%%%%%%%%%%%%%%%%%%%%%%%%%%
%%%%%%%%%%% figure
%%%%%%%%%%%%%%%%%%%%%%%%%%%%%%%%%%%%%%%%%%%%%%%%%%%%%%%%%%%%%%%%%%%%
\begin{figure*}[t]
    \begin{overpic}[width=0.48\columnwidth]{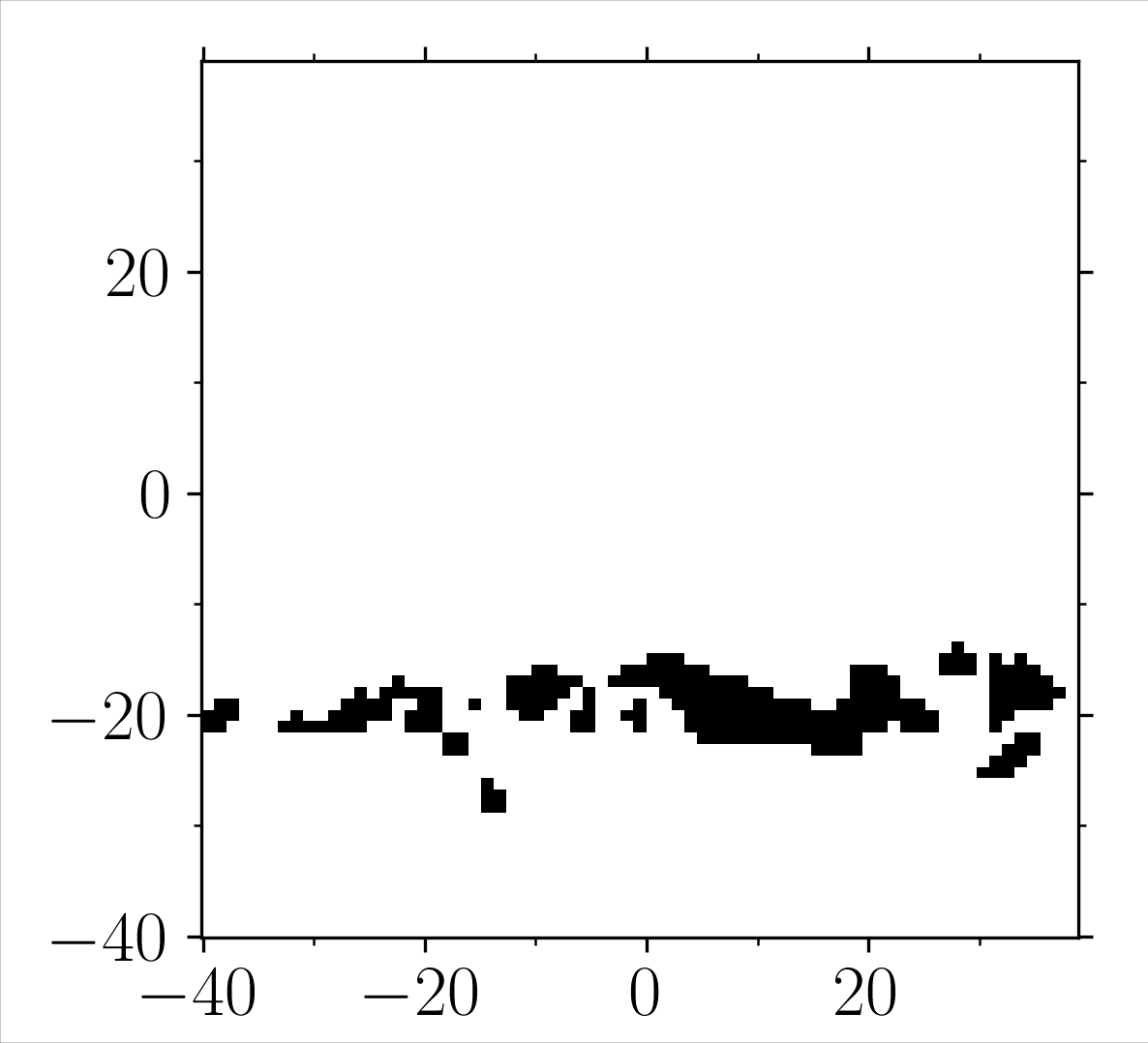}
        \LabelFig{19}{90}{$a)~\gamma_{xy} = 0.1$}
        \Labelxy{46}{-5}{0}{$x$(\r{A})}
        \Labelxy{-6}{44}{90}{$y$(\r{A})}
    \end{overpic}
    \begin{overpic}[width=0.49\columnwidth]{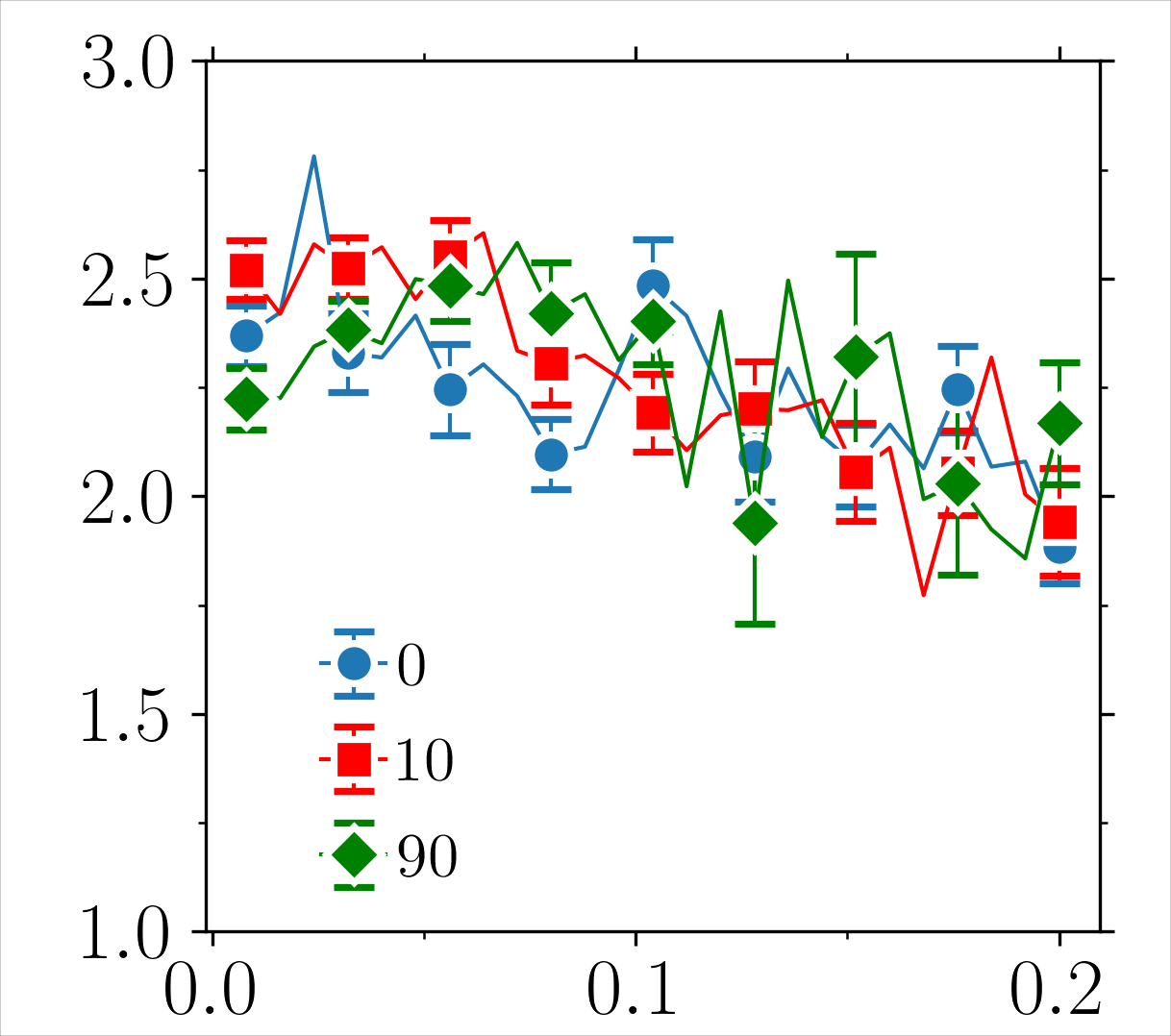}
         \LabelFig{19}{88}{$b)$}
        \Labelxy{50}{-5}{0}{$\gamma_{xy}$}
        \Labelxy{-2}{46}{90}{$d_f$}
        \Labelxy{20}{12}{90}{ $\scriptstyle t_\text{age}(\text{ps})$ }
    \end{overpic}
    % \vspace{-4pt}
    % 
    %
    \begin{overpic}[width=0.49\columnwidth]{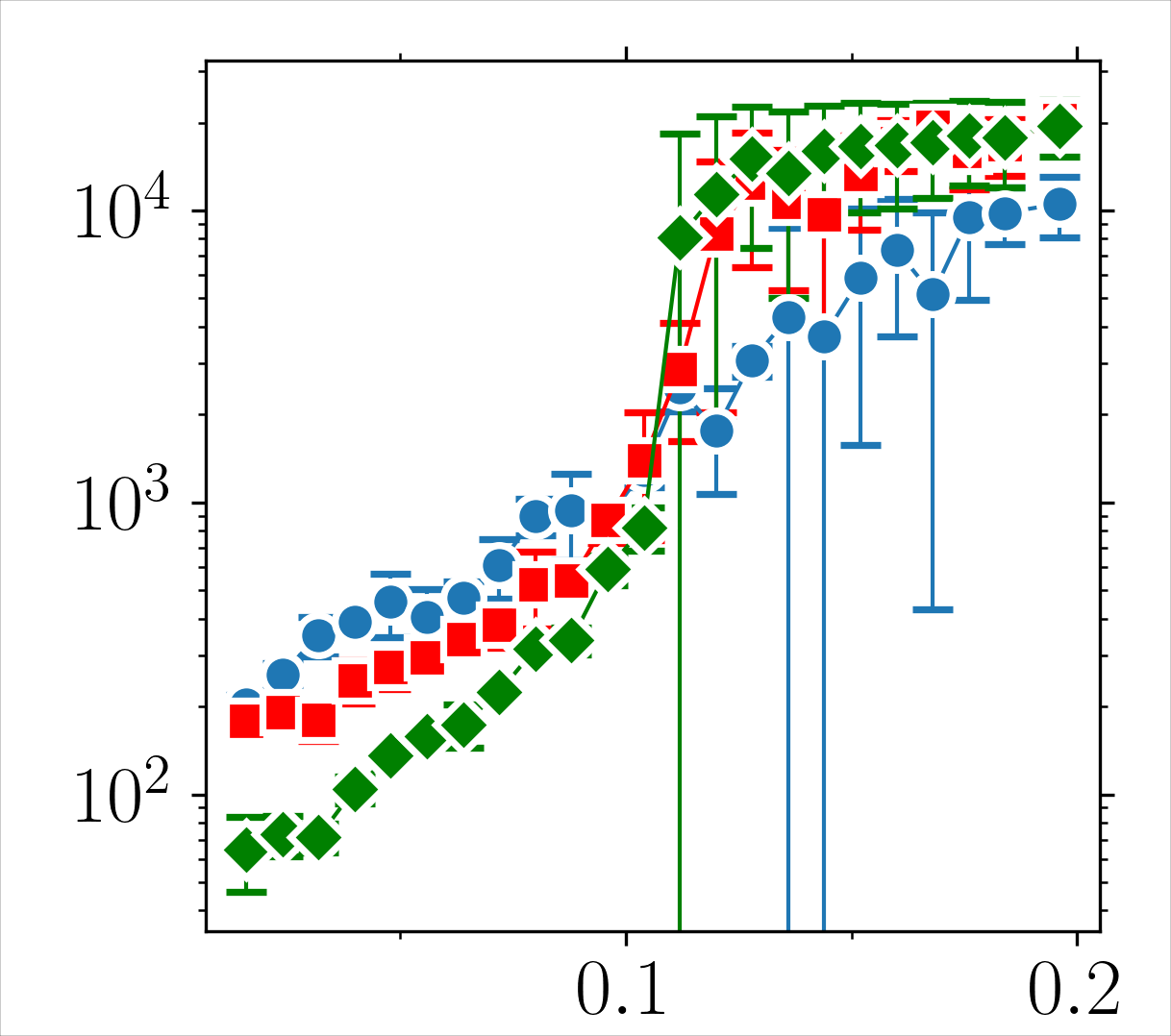}
         \LabelFig{19}{88}{$c)$}
        \Labelxy{50}{-5}{0}{$\gamma_{xy}$}
        \Labelxy{-2}{46}{90}{{$S$\tiny (\r{A}$^3$)}}
    \end{overpic}
    \begin{overpic}[width=0.49\columnwidth]{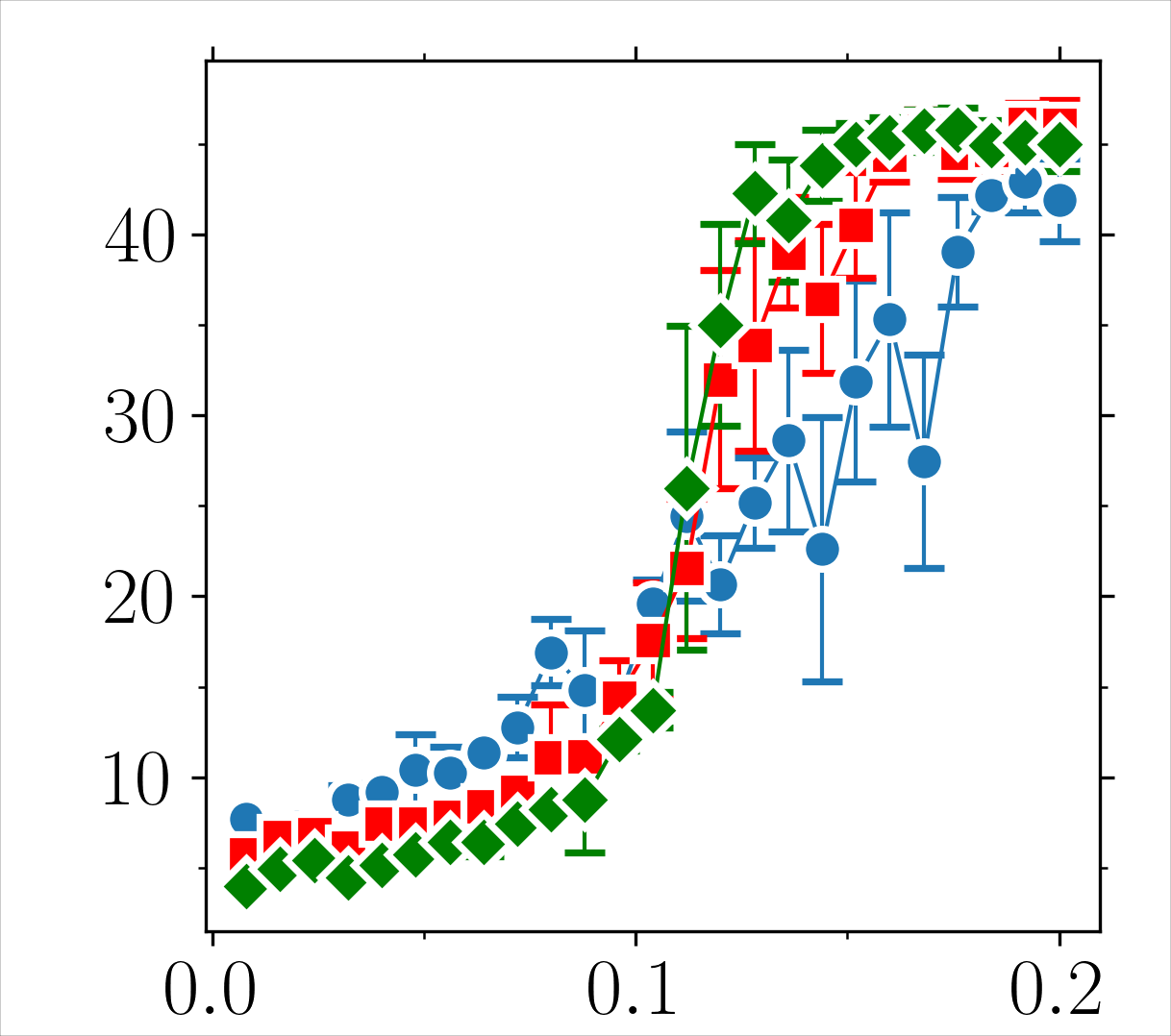}
         \LabelFig{19}{88}{$d)$}
        \Labelxy{50}{-5}{0}{$\gamma_{xy}$}
        \Labelxy{-2}{46}{90}{{$\xi$\tiny (\r{A})}}
    \end{overpic}
    %
%%%%%%%%%%%%%%%%%%%%%%%%%%%%%%%%%%%%%%%%%%%%%%%%%%%%%%%%%%%%%%%%%%%%
    \caption{\textbf{a}) Binary \dmin map at  $\gamma_{xy}=0.1$ corresponding to the annealed \glfive at \tage$=90$ ps. \textbf{b}) Fractal dimension $d_f$ \textbf{c}) Mean cluster size $S$ \textbf{d}) Correlation length $\xi$ plotted against strain $\gamma_{xy}$ at multiple age $t_\text{age}$. Here $x$ and $y$ denote flow and gradient directions, respectively. The binary maps show top $5\%$ sites with largest \dmin in black.}
    \label{fig:crltnGlassFive}
\end{figure*}

\subsection{Results}
Dynamics of individual atoms is quantified by $D^2_\text{min}$ as a measure of deviations from affine trajectories of atoms imposed by an external (homogeneous) shear \cite{falk1998dynamics}.
The $D^2_\text{min}$ analysis involves partitioning of the entire box into sub-volumes using a cubic grid of size $r_c=3.0$~\r{A}. %which is close to twice the radius associated with the first peak of $g(r)$.
We fit a linear model based on the reference positions $\vec{\mathring{r}}$ at zero strain, i.e $\vec{\mathcal{R}}(\vec{\mathring{r}})={\bf{F}} \vec{\mathring{r}}+\vec{F}_0$, to the current atomic positions $\vec{r}$ associated with $N_b$ atoms within a given volume by minimizing the squared loss function $\mathcal{L}({\bf{F}},
\vec{F}_0)=\sum_{i=1}^{N_b}|\vec{r}_i-\vec{\mathcal{R}}(\vec{\mathring{r}}_i)|^2$ to obtain $\{\hat{\bf{F}},\hat{\vec{F}}_0\}=\text{argmin}~\mathcal{L}({\bf{F}},
\vec{F}_0)$.
Here the second-ranked deformation tensor $\bf{F}$ and the (rigid) translation vector $\vec{F}_0$ correspond to an affine transformation.
The squared nonaffine displacements for atom $i$ is defined as $D^2_{i}=|\vec{r}_i-\hat{\vec{r}}_i|^2$ with $\hat{\vec{r}}_i$ being the predicted position $\hat{\vec{\mathcal{R}}}(\vec{\mathring{r}})={\hat{\bf{F}}} \vec{\mathring{r}}+\hat{\vec{F}}_0$ due to affine deformations.
%Squared nonaffine displacements $D^2_\text{min}$ maps and associated (binary) elasticity maps in sheared \glfive are displayed in Fig.~\ref{fig:map_d2min}(a-h) at different applied strains $\gamma_{xy}$.
%Negative-$\mu$ clusters tend to form percolating networks on approach to failure that somewhat correlate with the squared nonaffine displacements \dmin deformation maps.

Figure~\ref{fig:orderParametersGlassFive} plots the evolution of correlation coefficient $c_{XY}$, mean number density of atoms with icosahedral symmetries $\langle\rho^\text{ico}\rangle$, mean squared nonaffine displacements $\langle$\dmin\!\!$\rangle$, and scaled standard deviation std(\dmin\!)/$\langle$\dmin\!\!$\rangle$ associated with atoms' \dmin against applied strain $\gamma_{xy}$ in \glfive corresponding to different aging duration $t_\text{age}$.
Figure~\ref{fig:crltnGlassFive} displays fractal dimension $d_f$, mean cluster size $S$, and correlation length $\xi$ plotted against strain $\gamma_{xy}$ at multiple age $t_\text{age}$ corresponding to \glfive\!\!.
%%%%%%%%%%%%%%%%%%%%%%%%%%%%%%%%%%%%%%%%%%%%%%%%%%%%%%%%%%%%%%%%%%%%
%%%%%%%%%%% figure
%%%%%%%%%%%%%%%%%%%%%%%%%%%%%%%%%%%%%%%%%%%%%%%%%%%%%%%%%%%%%%%%%%%%
% \begin{figure}[t]
%     \begin{overpic}[width=0.5\columnwidth]{Figs2nd/rhoIco_strain_age.png}
%         \Labelxy{50}{-3}{0}{$\gamma_{xy}$}
% %        \Labelxy{-2}{32}{90}{$\langle$ \dmin $\rangle$ \tiny(\r{A}$^2$)}
%         \Labelxy{-5}{32}{90}{$\langle$\rhoico\!$\rangle$ \tiny(\r{A}$^{-3}$)}
%         \put(100,30){\includegraphics[width=0.07\columnwidth]{Figs2nd/rhoIco_strain_age_legend.png}}  
%         \Labelxy{100}{83}{0}{ $\scriptstyle t_\text{age}(\text{ps})$ }
%     \end{overpic}
%     \caption{Mean number density of atoms with icosahedral symmetries $\langle\rho^\text{ico}\rangle$ plotted against applied strain $\gamma_{xy}$ corresponding to different aging duration $t_\text{age}$. The error bars denote standard errors.}
%     \label{fig:rhoico}
% \end{figure}

\clearpage
\bibliography{references}